%% file: km_final.tex
\documentclass[twocolumn, times, trackchanges, tighten]{aastex63}

\usepackage{amsmath,amstext}
\usepackage[figure,figure*]{hypcap}
\usepackage{newtxmath} 

\received{2021 December 14}
\accepted{2022 February 2 }

\defcitealias{Worseck2019}{W19}
\defcitealias{Makan2021}{M21}
\defcitealias{Davies2017}{D17}

\begin{document}

\title{
\ion{He}{2} Ly$\boldsymbol{\alpha}$ Transmission Spikes and Absorption Troughs 
in Eight High-resolution Spectra Probing the End of \ion{He}{2} Reionization
}

\shorttitle{\ion{He}{2} Ly$\alpha$ Transmission Spikes and Absorption Troughs}
\shortauthors{Makan et al.}

\correspondingauthor{Kirill Makan}
\email{kmakan@astro.physik.uni-potsdam.de}

\author[0000-0003-3157-1191]{Kirill Makan}
\affiliation{Institut f\"ur Physik und Astronomie, Universit\"at Potsdam, Karl-Liebknecht-Str. 24/25, 
D-14476 Potsdam, Germany}

\author[0000-0003-0960-3580]{G\'abor Worseck}
\affiliation{Institut f\"ur Physik und Astronomie, Universit\"at Potsdam, Karl-Liebknecht-Str. 24/25, 
D-14476 Potsdam, Germany}

\author[0000-0003-0821-3644]{Frederick B. Davies}
\affiliation{Max-Planck-Institut f\"ur Astronomie, K\"onigstuhl 17, D-69117 Heidelberg, Germany}

\author[0000-0002-7054-4332]{Joseph F. Hennawi}
\affiliation{Department of Physics, University of California, Santa Barbara, CA 93106-9530, USA}
\affiliation{Leiden Observatory, Leiden University, Niels Bohrweg 2, NL-2333 CA Leiden, the Netherlands}

\author[0000-0002-7738-6875]{J. Xavier Prochaska}
\affiliation{University of California Observatories, Lick Observatory, University of California, Santa Cruz, CA 95064, USA}
\affiliation{Kavli Institute for the Physics and Mathematics of the Universe (WPI), UTIAS, The University of Tokyo, Kashiwa, Chiba 277-8583, Japan}

\author[0000-0002-1188-1435]{Philipp Richter}
\affiliation{Institut f\"ur Physik und Astronomie, Universit\"at Potsdam, Karl-Liebknecht-Str. 24/25, 
	D-14476 Potsdam, Germany}

\begin{abstract}
We present statistics of \ion{He}{2} Ly$\alpha$ transmission 
spikes and large-scale absorption troughs using archival 
high-resolution ($R=\lambda /\Delta \lambda \simeq 12,500$--$18,000$) 
far-UV spectra of eight \ion{He}{2}-transparent quasars obtained 
with the Cosmic Origins Spectrograph on the Hubble Space Telescope. 
The sample covers the redshift range $2.5 \lesssim z \lesssim 3.8$, 
thereby probing the rapidly evolving \ion{He}{2} absorption at the 
end of \ion{He}{2} reionization epoch. The measured lengths of the 
troughs decrease dramatically from $L\gtrsim 100\mathrm{\,cMpc}$ at 
$z > 3$ to $L\simeq30\mathrm{\,cMpc}$ at $z \sim 2.7$, signaling a 
significant progression of \ion{He}{2} reionization at these 
redshifts. Furthermore, unexpectedly long  $L\sim65\mathrm{\,cMpc}$ 
troughs detected at $z\simeq2.9$ suggest that the UV background 
fluctuates at larger scales than predicted by current models. 
By comparing the measured incidence of transmission spikes 
to predictions from forward-modeled mock spectra created from the 
outputs of a $(146\mathrm{\,cMpc})^{3}$ optically thin \texttt{Nyx} 
hydrodynamical simulation employing different UV background models, 
we infer the redshift evolution of the \ion{He}{2} photoionization 
rate $\Gamma_\mathrm{He\,II}(z)$. The photoionization rate decreases 
with increasing redshift from $\simeq 4.6\times 10^{-15}\mathrm{\,s^{-1}}$ 
at $z\simeq 2.6$ to $\simeq 1.2 \times 10^{-15}\mathrm{\,s^{-1}}$ 
at $z\simeq3.2$, in agreement with previous inferences from 
the \ion{He}{2} effective optical depth, and following expected trends 
of current models of a fluctuating \ion{He}{2}-ionizing background. 

\end{abstract}


\section{Introduction} \label{sec:intro}

The reionization of intergalactic primordial gas completed 
separately for hydrogen and helium. For hydrogen, several 
observational tracers, such as the cosmic microwave background 
\citep{PlanckCollab2018}, Ly$\alpha$ emitters 
\citep[e.g.,][]{Mason2018, Hu2019} and Lyman series 
absorption \citep[e.g.,][]{Fan2006}, suggest that hydrogen 
reionization ended at $z \sim 6$ or even somewhat later 
\citep[e.g.,][]{Becker2015, Kulkarni2019, Choudhury2021}. 
Due to their similar ionization potential, hydrogen and 
\ion{He}{1} were reionized by soft UV photons that were likely 
emitted by  star-forming galaxies 
\citep[e.g.,][]{Madau1999, FaucherGiguere2008, Eide2020}. 
\ion{He}{2} reionization, however, requires hard UV photons 
($E \ge 54.4\mathrm{\,eV}$) from quasars and was completed 
later at $z \sim 3$ \citep[e.g.,][]{MadauMeiksin1994, Fardal1998,  
Miralda-Escude2000, Sokasian2002, FurlanettoOh2008, McQuinn2009b, 
Compostella2013, Compostella2014, LaPlante2017, Kulkarni2019}. 
The progression of \ion{He}{2} reionization determines the 
time evolution and spatial variance in the intergalactic 
UV radiation field 
\citep[e.g.,][]{FurlanettoDixon2010,Davies2017,Meiksin2020}, 
and thus, is a crucial probe of observational cosmology.

Depending on the employed parameters such as the quasar 
luminosity function and spectral energy distribution, 
the current hydrodynamical simulations result in a broad 
range of \ion{He}{2} reionization histories ending at 
$2.3 \lesssim z \lesssim 3.4$
\citep{Compostella2013, Compostella2014,  LaPlante2017}. 
Primarily, these simulations were used to predict the redshift 
evolution of the IGM temperature as inferred from the 
\ion{H}{1} Ly$\alpha$ forest 
\citep[e.g.,][]{Boera2014, Hiss2018, Walther2019}.

In principle, \ion{He}{2} Lyman series absorption can be observed 
in the far-UV (FUV) against $z > 2$ quasars 
\citep[e.g.,][]{Miralda-Escude1993,MadauMeiksin1994,Jakobsen1994}. 
In practice, however, cumulative intergalactic \ion{H}{1} Lyman 
continuum absorption and the rapidly declining $z>3$ quasar 
luminosity function strongly limit the number of quasars with FUV 
emission at the \ion{He}{2} Ly$\alpha$ rest-frame wavelength 
$\lambda_\alpha=303.7822$\,\AA\ 
\citep{MollerJakobsen1990,PicardJakobsen1993,WorseckProchaska2011}. 
Statistical \ion{He}{2} absorption studies toward $>20$ quasars 
were enabled only recently with pre-selection of FUV-bright quasars 
from Galaxy Evolution Explorer \citep[GALEX;][]{Morrissey2007} 
photometry \citep{Syphers2009a,Syphers2009b,WorseckProchaska2011} 
and spectroscopic follow-up with the FUV-sensitive Cosmic Origins 
Spectrograph (COS) onboard the Hubble Space Telescope 
\citep[HST,][]{Worseck2011,Syphers2012,Worseck2016,Worseck2019}.

One of the simplest \ion{He}{2} absorption statistics is the 
\ion{He}{2} Ly$\alpha$ effective optical depth 
$\tau_\mathrm{eff} = -\ln \langle e^{-\tau _\alpha} \rangle _{\Delta z}$ 
over a predefined redshift interval $\Delta z$. 
In the current sample of 25 \ion{He}{2} sight lines, the variance of 
$\tau_\mathrm{eff}$ is in agreement with IGM density fluctuations in 
a uniform UV background at $z\lesssim 2.7$, indicating 
the end of the \ion{He}{2} reionization epoch 
\citep[][hereafter \citetalias{Worseck2019}]{Worseck2019}.
However, at higher redshifts ($2.7 \lesssim z \lesssim 3.2$), 
the much larger sightline-to-sightline variance of $\tau_\mathrm{eff}$ 
requires a fluctuating UV background \citep{Davies2017} at still 
very low \ion{He}{2} fractions of $\sim 1$\% 
(\citealt{Worseck2016}; \citealt{Davies2017}; \citetalias{Worseck2019}). 
Because the $\tau_\mathrm{eff}$ statistic is readily applicable to 
the available statistically useful sample of low-quality low-resolution 
\ion{He}{2} spectra it remains the most direct observational probe 
of the \ion{He}{2} reionization epoch, analogous to \ion{H}{1} 
at $z\gtrsim 5.5$ \citep[e.g.,][]{Fan2006, Eilers2019, Bosman2021}.

Alternative approaches, such as statistics of transmission 
spikes \citep{Gallerani2006, Gallerani2008, Gnedin2017, 
Barnett2017, Chardin2018, Garaldi2019, Gaikwad2020, Yang2020} 
and troughs between them 
\citep{SongailaCowie2002, Paschos2005, Fan2006, Gallerani2006, 
Gallerani2008, Gnedin2017, Zhu2021}, have already been used 
to probe the end of the \ion{H}{1} reionization epoch. 
The same statistics may also distinguish between different 
\ion{He}{2} reionization models \citep{Compostella2013}.
However, so far they have been applied to only two $z>3.1$ 
\ion{He}{2} sightlines with the required high-resolution 
data (\citealt{Makan2021}; hereafter \citetalias{Makan2021}).

Here we present statistics of \ion{He}{2} Ly$\alpha$ transmission 
spikes and troughs in the high-resolution ($R = 12,500$--$18,000$) 
HST/COS spectra of eight \ion{He}{2}-transparent quasars 
covering $2.5 \lesssim z \lesssim 3.8$. By employing the automated 
transmission spike measurement technique presented in 
\citepalias{Makan2021}, we aim to probe the end of the \ion{He}{2} 
reionization epoch, providing an alternative approach to inferences 
from $\tau_\mathrm{eff}$. This paper is structured as follows. 
In Section~\ref{sec:obs_data_reduction}, we describe our sample 
and our custom data reduction. In Section~\ref{sec:methods}, 
we explain and apply our measurement technique for transmission 
spikes and troughs. We compare the measured incidence of 
transmission spikes to predictions from photoionization models to 
infer the redshift-dependent \ion{He}{2} photoionization rate 
$\Gamma_\mathrm{He\,II}(z)$ in 
Section~\ref{sec:trans_features_in_mock_spectra}. 
Finally, we summarize in Section~\ref{sec:summary}. 
We use a flat cold dark matter cosmology with dimensionless Hubble 
constant $h = 0.685$ ($H_{0} = 100h\mathrm{\,km\,s^{-1}\,Mpc^{-1}}$) 
and density parameters 
$(\Omega _{\mathrm{m}}, \Omega _{\mathrm{b}}, \Omega _{\mathrm{\Lambda}}) = (0.3, 0.047, 0.7)$, 
consistent with \citet{PlanckCollab2018}.

\input{tab1}

 \section{High resolution {HST/COS spectra of \ion{He}{2}-transparent quasars}}
 \label{sec:obs_data_reduction}

 \subsection{Archival HST/COS Data }

Our sample of \ion{He}{2}-transparent quasars with high-resolution 
HST/COS \citep{Green2012} G130M spectra includes eight objects with 
$z_{\mathrm{em}} = 2.74$--$3.81$ (Table~\ref{tab:quasar_sample}). 
The two quasars at $z_\mathrm{em} > 3.5$ were recently observed in 
HST Cycle 25 Program 15356 at COS Lifetime Position (LP) 4 and 
studied in \citetalias{Makan2021}. Exposures taken at several 
central wavelengths ensured a gap-free coverage from 1067\,\AA\ to 
1473\,\AA\ at a wavelength-dependent resolving power of 
$R =  14,000$ at 1250\,\AA. We complemented these data with 
archival HST high-resolution spectra of six lower-redshift 
\ion{He}{2}-transparent quasars.
HE\,2347$-$4342, one of the FUV-brightest \ion{He}{2}-transparent 
quasars, was observed in Program 11528 at LP~1 providing 
$R\simeq 18,000$ at 1250\,\AA\ \citep{Shull2010}. 
It was later reobserved in Program 13301 at LP~2 
with the short-wavelength setup centered at 1222\,\AA\, 
along with HS\,1700$+$6416  \citep{ShullDanforth2020}. 
The same 1222\,\AA\ setup was used for the
quasars HS\,1024$+$1849, Q\,1602$+$576 and HS\,0911$+$4809 in
Program 12816 (PI Syphers).
Finally, Q\,0302$-$003 was observed with COS
in Program 12033 at LP~1 with the central wavelength settings
1291\,\AA\ and 1318\,\AA\  \citep{SyphersShull2014}.

\subsection{Data Reduction}
\label{sec:data_reduction}
All data were reduced following \citetalias{Makan2021} by using
\texttt{CALCOS} v3.3.9 along with our improved FUV data reduction 
pipeline \texttt{FaintCOS}
\footnote{\href{https://github.com/kimakan/FaintCOS}
{https://github.com/kimakan/FaintCOS}} v1.0, which has been 
extensively described and tested in \citetalias{Makan2021}. 
\texttt{FaintCOS} is the only publicly available pipeline that 
enables accurate dark current subtraction and 
co-addition of subexposures of faint objects taken in the Poisson 
regime at the sensitivity limit of HST/COS. 
We adjusted the reduction parameters in \texttt{FaintCOS} for 
every object individually to account, among other things, 
for the detector degradation and changing observing conditions.

For some quasars, the data were obtained at different epochs 
and thus require a flux scaling to account for potential 
quasar variability and varying sensitivity of the COS 
detector. For instance, the flux of HE\,2347$-$4342 decreased 
by $\sim10$\,\% between 2009 November and 2014 August. 
Furthermore, the data set from 2014 August of this particular 
sight line showed an abrupt change in continuum flux between 
FUV detector segment A and B owing to the changed high-voltage 
settings on the latter. We used the low resolution G140L 
HST/COS data (Program 13301) to rescale the individual exposures. 
On the other hand, the rescaling for HS\,1700$+$6416 was not required 
because the relative flux difference between 2014 April and 
2014 December, although taken at different high-voltage levels, 
remained $<2\%$ which is in agreement with the
HST/COS absolute flux 
calibration accuracy of $5\%$.

The single exposure spectra were co-added and rebinned in count 
space to Nyquist-sampled spectra with $0.04\mathrm{\AA\,pixel^{-1}}$ 
($0.03\mathrm{\AA\,pixel^{-1}}$ for Q0302$-$003), 
preserving the Poisson distributed counts. 
Q0302$-$003 has a different binning due to the 
higher spectral resolution $R = 18,000$ at 1250\,\AA\, 
($R = 12,500$--$14,500$ for other data). 
A comparison of the individual exposures 
(and visits for the faintest objects)  
did not reveal any systematic wavelength calibration differences, 
preserving the spectral resolution for \ion{He}{2} transmission spikes.
The flux calibration curves and flat fields of 
individual exposures determined by \texttt{CALCOS} were combined 
using the pixel exposure time weighted average.

We suppressed the contamination by geocoronal emission lines 
(\ion{N}{1}~$\lambda 1134$, \ion{N}{1}~$\lambda 1168$, 
\ion{N}{1}~$\lambda 1200$, \ion{H}{1}~$\lambda 1216$, 
\ion{N}{1}~$\lambda 1243$, \ion{O}{1}~$\lambda 1304$, 
\ion{O}{1}~$\lambda 1356$) by replacing the affected spectral 
regions with the data taken during orbital night 
(Sun altitude $\leqslant 0^{\circ}$). We were able to remove 
all geocoronal emission, except Ly$\alpha$, along 
most of our sight lines in the \ion{He}{2} Ly$\alpha$ 
absorption region of the spectra. Q\,0302$-$003 and 
HS\,0911$+$4809 contain obvious residuals of 
\ion{N}{1}~$\lambda 1134$, \ion{N}{1}~$\lambda 1200$, 
\ion{O}{1}~$\lambda 1304$. Spectral regions with residual 
contamination are excluded from the analysis. 
All ISM $\mathrm{H_2}$ fluorescence lines are considered 
to be negligible, because even 
the strongest lines are not present in the predominantly 
saturated spectra of HE2QS\,J2311$-$1417 and 
HE2QS\,J1630$+$0435 \citepalias{Makan2021}. 
The weak diffuse UV sky background $f_{\mathrm{sky}}$ was 
subtracted by using GALEX data \citep{Murthy2014} following the 
procedure by \citet{Worseck2016}.

We calculated the asymmetric statistical errors of the 
background-subtracted flux in the Poisson regime at 68.27\% 
confidence level using the method by \citet{FeldmanCousins1998}. 
The resulting signal-to-noise ratio (S/N) is 3--28 per pixel 
near \ion{He}{2} Ly$\alpha$ in the quasar continuum. 
The S/N varies greatly between the sight lines owing 
to the different exposure times and quasar brightness. 
The reduced spectra were corrected for Galactic extinction 
by applying the extinction curve $A_\lambda$ from \citet{Cardelli1989}. 
For the correction, we used the line-of-sight selective extinction 
from \citet{Schlegel1998} and assumed the Galactic average ratio 
between the total $V$ band extinction and selective extinction
$R_{V}$=3.1.

As in \citetalias{Worseck2019} and \citetalias{Makan2021}, 
we used low-resolution COS and STIS G140L spectra to 
define quasar power-law continua 
$E_{\mathrm{\lambda}} \propto \lambda ^{\alpha}$. 
For this, absorption-free regions in the quasar continuum were 
manually selected and fitted using the maximum likelihood method. 
Additionally, we determined the redshifts of partial Lyman limit 
systems through their Lyman series absorption lines 
and added their \ion{H}{1} column densities as free parameters 
to the power-law fit. 
The resulting continua were scaled to the high-resolution 
G130M data using the overlapping quasar continua redward of \ion{He}{2} 
Ly$\alpha$. The results are listed in Table~\ref{tab:quasar_sample} 
along with the identified partial Lyman limit systems.

\begin{figure*}
	\includegraphics[width=\textwidth]{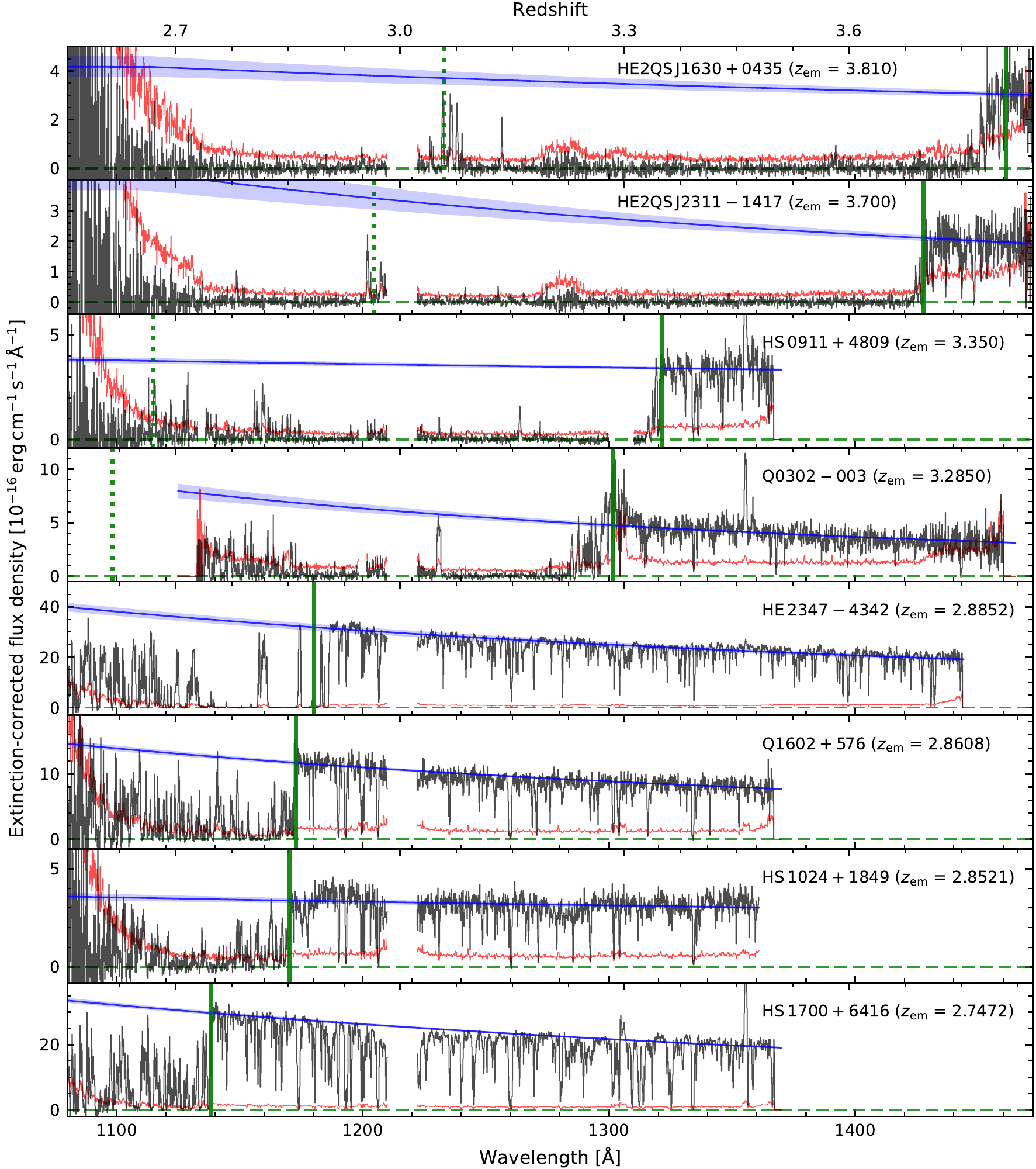}
	\caption{HST/COS G130M spectra (black) and
	their corresponding Poisson single-sided upper error 
	($84.13$\% confidence level, red) 
	of eight \ion{He}{2}-transparent quasars sampled at 
	0.12\,\AA\,pixel$^{-1}$ for visualization. The strong 
	geocoronal \ion{H}{1} Ly$\alpha$ emission line in all 
	spectra, as well as the geocoronal \ion{O}{1} in the 
	spectrum of HS\,0911$+$4809, is omitted. The green lines 
	indicate \ion{He}{2} Ly$\alpha$ at the redshift of
	the quasars with the quasar continuum redwards of it. 
	The reduced flux at $\sim$1205\,\AA\, and $\sim$1225\,\AA\,  
	in the quasar continua are the damping wings of the 
	Galactic \ion{H}{1} Ly$\alpha$ absorption. The green dotted 
	lines mark \ion{He}{2} Ly$\beta$. The fitted continua 
	from HST low-resolution spectra and the corresponding 
	1$\sigma$ error are shown in blue.
	}
	\label{fig:spectra_overview}
\end{figure*}

\subsection{General Overview of the Spectra}
\label{sec:general_description}

Figure~\ref{fig:spectra_overview} shows the high-resolution
HST/COS G130M spectra of our eight quasars (undersampled for 
visualization) along with their respective fitted and scaled 
continua. The substantial decrease in S/N at $\lambda<1120$\,\AA\ 
is due to the decreasing sensitivity of the detector at these 
wavelengths, while the extended scatter at $1280$\,\AA\, 
in the spectra of the two $z_\mathrm{em} > 3.5$ quasars is 
due to the short exposure time resulting from partial overlap 
of multiple central wavelength settings.

Blueward of \ion{He}{2} Ly$\alpha$ in the quasar rest frame 
we observe predominant intergalactic \ion{He}{2} Ly$\alpha$ 
absorption. Enhanced ionization around the background 
quasar is visible as excess transmission in its proximity zone 
\citep{Khrykin2019, Worseck2021}. 
Noticeably, the \ion{He}{2} Ly$\alpha$ absorption
is predominantly saturated at $z > 3$, but showing occasional 
transmission spikes (e.g., at $z\sim 3.17$ toward HS\,0911$+$4809 
and at $z\sim 3.58$ toward HE2QSJ1630$+$0435). In contrast at 
lower redshifts (e.g., $z \lesssim 2.7$ toward HE\,2347$-$4342), 
there is already a lot of small-scale structure that resembles 
an emerging \ion{He}{2} Ly$\alpha$ forest. Three of the spectra 
also cover intergalactic \ion{He}{2} Ly$\beta$, discussed for 
the two $z>3.5$ quasars in \citetalias{Makan2021}.

\section{Observed \ion{He}{2} Transmission Features}
\label{sec:methods}

\subsection{Measurement Technique}

Transmission spikes were detected and fitted with our fully 
automated spike finding code presented in \citetalias{Makan2021}. 
The algorithm decomposes the spectrum into a series of Gaussian 
transmission spikes. Obviously, absorption features cannot be 
modeled as emission features, but our approach allows for a 
fast one-to-one analysis of predominantly saturated spectra. 
Here we also perform measurements of the 
troughs to obtain an additional statistic of \ion{He}{2} 
reionization \citep[e.g.,][]{Compostella2013}.

In short, we used the Poisson probability
\begin{equation}
    P(>N|B) = 1 - \sum _{k=0}^{N} \frac{B^{k}e^{-B}}{k!}
\end{equation}
for $N$ Poisson counts given the background $B$
as a 9-pixel (0.27\,\AA\ -- 0.36\,\AA) 
running average to define spectral regions with significant 
($P(>N|B) < 0.0014$, $3\sigma$) transmission. In order to 
reliably fit Gaussian profiles to the individual transmission 
spikes even 
in low signal-to-noise spectra at $z>3$, 
we required that a significantly transmitting region 
must have at least nine consecutive $P < 0.0014$ pixels. 
After smoothing of the transmission with a 
Gaussian filter ($\sigma_\mathrm{f} = 0.1$\,\AA), individual spikes 
were found by counting all transmission maxima in these spectral 
regions. Following \citetalias{Makan2021}, our primary 
statistic for transmission spikes is their incidence, 
i.e., number of maxima in predefined redshift bins. 
All our measurements were carried out in the valid spectral 
regions which exclude the \ion{He}{2} Ly$\alpha$ proximity 
zone \citep{Khrykin2019, Worseck2021}, the overlapping 
\ion{He}{2} Ly$\beta$ forest, regions contaminated by 
geocoronal emission lines, and regions of very low 
sensitivity resulting in $P > 0.0014$ even at full 
transmission ($\tau_\alpha=0$).

We defined spectral regions between transmission spikes as 
troughs. These regions do not contain significant 
small-scale transmission given our data quality. For every trough, 
we measured the length $L$ and the effective optical depth 
$\tau _{\mathrm{eff}}$. 
Our definition of the troughs depends on the 9-pixel averaged 
$P(>N|B)$, and thus, the trough lengths are measured at 
$\sim 1$\,cMpc accuracy. The troughs are truncated by transmitting or 
invalid spectral regions, e.g., geocoronal emission, which we take 
into account in the following analysis. The \ion{He}{2} Ly$\alpha$ 
effective optical depth is 
$\tau _{\mathrm{eff}} = -\ln \langle f_{\lambda} / E_{\lambda} \rangle_L$, 
where $\langle \rangle_L$ is the average over the trough length $L$. 
In practice, we used the maximum likelihood method in the Poisson 
regime to measure $\tau _{\mathrm{eff}}$ (see \citetalias{Worseck2019} 
or \citetalias{Makan2021} for details).
We caution that due to our definition, troughs do not have a lower 
limit on $\tau _{\mathrm{eff}}$. Thus, $\tau _{\mathrm{eff}}$ can be 
measured in troughs at high significance even if small-scale spikes 
remain undetectable. In order to exclude possible individual strong 
\ion{He}{2} Ly$\alpha$ absorption lines 
(column density $N _{\mathrm{He\,II}} \gtrsim 10^{19} \mathrm{\,cm^{-2}}$) 
from our trough measurements, we only consider lengths 
$L \geqslant 10$\,cMpc. We emphasize that our trough measurement 
method depends on our ability to detect transmission spikes via 
$P(>N|B)$ and thus, the low S/N spectra will generally contain 
more and longer troughs. Therefore, physical inferences require 
detailed forward-modeling (Section~\ref{sec:mock_spectra}).

\begin{figure*}\label{fig:low_z_spectra}
	\includegraphics[width=\textwidth]{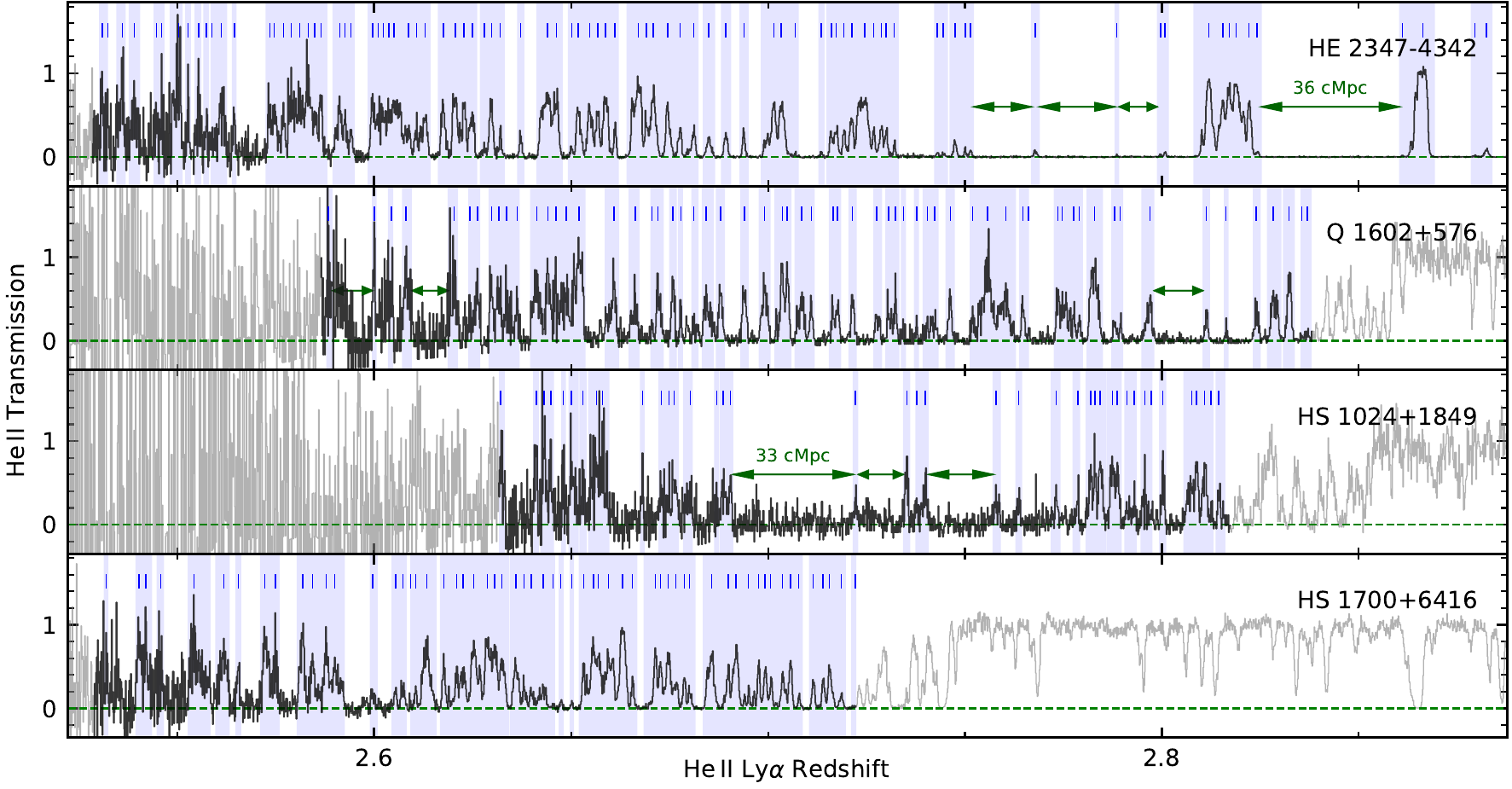}
	\caption{Detected \ion{He}{2} Ly$\alpha$ transmission 
	spikes and troughs at the original spectral binning of 0.04\,\AA 
	toward the four $z_{\mathrm{em}} < 2.9$ quasars. The spectral regions plotted 
	in gray are omitted from the analysis due to the low sensitivity at
	the short-wavelength end of the spectrum, and due to 
	the onset of the \ion{He}{2} Ly$\alpha$ proximity zone at the 
	long-wavelength end. Blue-shaded areas indicate spectral regions 
	with significant transmission ($P (>N|B)<0.0014$) in at least 
	9 consecutive pixels. The blue vertical lines mark individual 
	transmission spikes. The troughs (regions between 
	blue shaded areas with length $L \geqslant 10$\,cMpc) are indicated 
	with arrows.
	}	
\end{figure*}

\subsection{Observational Results}
\label{sec:obs_results}

\begin{figure*}\label{fig:mid_z_spectra}
	\includegraphics[width=\textwidth]{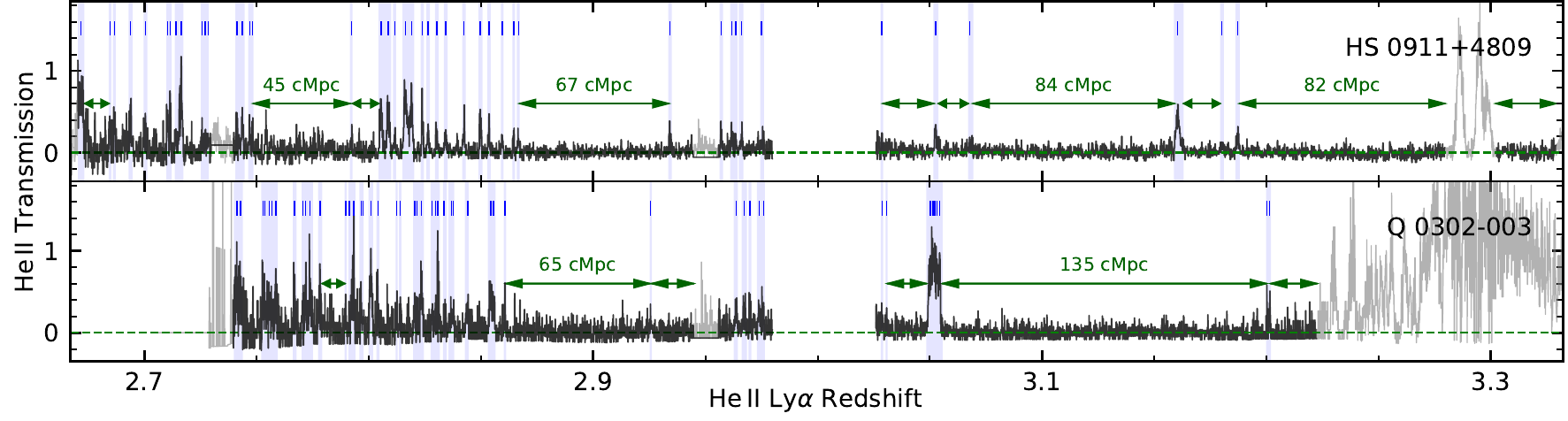}
	\caption{Similar to Figure~\ref{fig:low_z_spectra}, but for 
	sight lines at $z_{\mathrm{em}}=3.2$--$3.5$.
	Spectral regions with geocoronal Ly$\alpha$ $\lambda 1216$, 
	\ion{O}{1} $\lambda 1304$, \ion{N}{1} $\lambda 1200$ and 
	\ion{N}{1} $\lambda 1134$ emission 
	are omitted from the analysis. 
	}
\end{figure*}

\begin{figure*}\label{fig:high_z_spectra}
	\includegraphics[width=\textwidth]{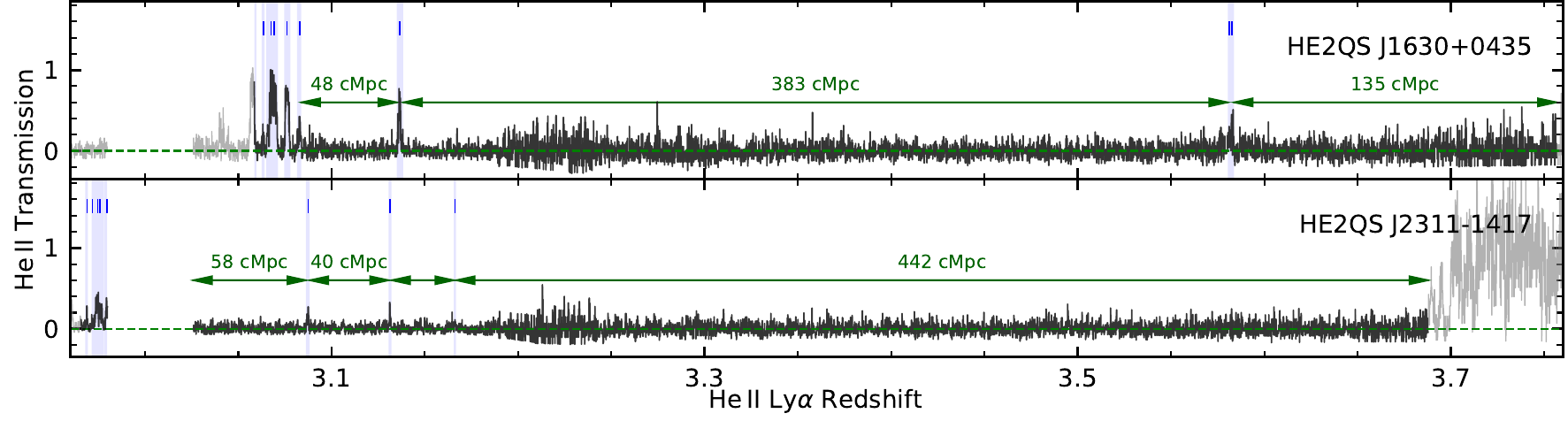}
	\caption{Similar to Figures \ref{fig:low_z_spectra} and 
	\ref{fig:mid_z_spectra}, but for the two  sight lines at 
	$z_{\mathrm{em}} > 3.5$. Spectral regions at the short-wavelength 
	end are omitted due to \ion{He}{2} Ly$\beta$ absorption. }
\end{figure*}

\begin{figure}\label{fig:spike_details}
	\includegraphics[width=\linewidth]{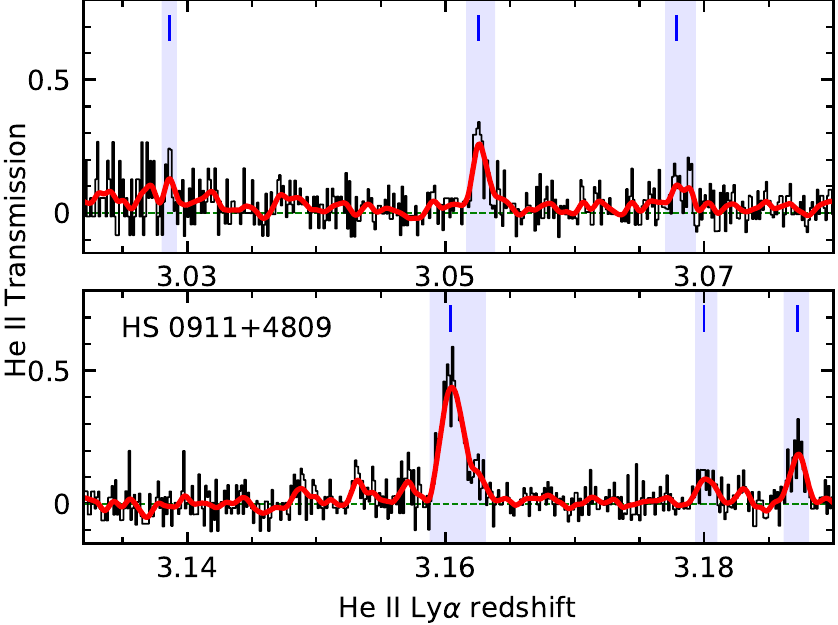}
	\caption{Similar to Figures \ref{fig:low_z_spectra}, 
	\ref{fig:mid_z_spectra} and \ref{fig:high_z_spectra}, but 
	zoomed-in on the $z > 3$ transmission spikes toward 
	HS\,0911$+$4809. The red line shows the transmission smoothed 
	with a Gaussian filter ($\sigma _f = 0.1$\,\AA).}
\end{figure}

Figures~\ref{fig:low_z_spectra}, \ref{fig:mid_z_spectra} and 
\ref{fig:high_z_spectra} visualize the results of our spike and trough 
measurements in the valid spectral regions. 
Additionally as an example, Figure~\ref{fig:spike_details} 
shows the $z>3$ transmission spikes toward HS\,0911$+$4809 in 
more detail. 
Table~\ref{tab:troughs} lists all measured troughs with their 
redshift ranges, lengths and effective optical depths. 
As expected, all four sightlines probing
$z < 2.7$ show an emerging \ion{He}{2} forest. The \ion{He}{2} Ly$\alpha$ 
transmission rarely reaches unity, indicating strong \ion{He}{2} 
absorption from underdense regions in the IGM that hampers fits to 
individual absorption lines based on the coeval \ion{H}{1} Ly$\alpha$ 
forest \citep{Reimers1997,Kriss2001, FechnerReimers2007, 
McQuinnWorseck2014, SyphersShull2014}. Only four short troughs 
($L \lesssim 30$\,cMpc) are detected at $z < 2.72$, exclusively 
in the low S/N spectra and therefore subject to data quality.

At $2.72<z<3$ the \ion{He}{2} Ly$\alpha$ absorption becomes patchy, 
i.e.\ there are alternating regions with clustered transmission spikes 
separated by longer troughs. This patchy \ion{He}{2} absorption is most 
prominent toward HE\,2347$-$4342 \citep{Reimers1997}, but persists to 
$z\sim 2.9$--3 toward HS\,0911$+$4809 and Q\,0302$-$003, shown here for 
the first time (Fig.~\ref{fig:mid_z_spectra}). The frequency of 
\ion{He}{2} Ly$\alpha$ transmission spikes dramatically decreases toward 
higher redshifts. At $3\lesssim z\lesssim 3.3$ we measure occasional 
and often isolated transmission spikes separated by $L\gtrsim 40$\,cMpc 
long troughs in four sightlines (Figs.~\ref{fig:mid_z_spectra} and 
\ref{fig:high_z_spectra}). The trough lengths appear to increase with 
increasing redshift as expected due to the higher IGM \ion{He}{2} 
fraction before the completion \ion{He}{2} reionization 
\citep[e.g.,][]{Compostella2013}. The longest detected \ion{He}{2} 
Ly$\alpha$ troughs (442\,cMpc toward HE2QS\,J2311$-$1417 and 383\,cMpc 
toward HE2QS\,J1630$+$0435) were already briefly mentioned in 
\citetalias{Makan2021}.

At $z>3$, the rare \ion{He}{2} transmission spikes typically arise 
from highly underdense regions of the IGM \citep{McQuinn2009} or 
highly ionized quasar proximity zones. Indeed, some of the spikes 
are very close to the background quasar proximity zones 
(e.g., Q\,1602$+$576 and Q\,0302$-$003), the sizes of which are 
likely underestimated due to their simple observational definition 
\citep{Khrykin2016}. Nevertheless, the inclusion of small regions 
with likely non-negligible quasar influence does not strongly affect 
our analysis. Likewise, some of the transmission spikes might be 
caused by the transverse proximity effect of foreground quasars 
\citep{Schmidt2017, Schmidt2018}. However, only one sight line 
in our sample (Q\,0302$+$003) has confirmed foreground quasars that 
are likely associated with the transmission feature at $z\sim 3.05$ 
\citep{Jakobsen2003, Schmidt2017}.

Our measurements disregard the mostly weak absorption from the 
low-redshift \ion{H}{1} Ly$\alpha$ forest overlapping with $z>3$ 
\ion{He}{2} Ly$\alpha$ absorption. This is justified, because it is 
very unlikely that rare saturated \ion{H}{1} absorption lines will 
black out rare \ion{He}{2} transmission spikes\footnote{At redshifts 
$z_\mathrm{H\,I} < 0.2$, corresponding to $z=3.0$--3.8 for \ion{He}{2}, 
there are on average $\simeq 5$ \ion{H}{1} absorbers with 
$\log N_\mathrm{H\,I} = 14$--17 \citep{Kim2021} per line of sight. 
Each of these absorbers has an optical depth at the Ly$\alpha$ line 
center $\tau _\mathrm{H\,I} > 1.7$ and full width at half maximum 
$\sim 0.3$\AA\ after accounting for the COS line-spread-function. 
The probability that at least one transmission spike overlaps with 
a Ly$\alpha$ absorption line is $\sim 0.6 - 6\,\%$ depending on the 
number of spikes (1--10).}. 
Moreover, our ability to detect transmission spikes depends on the S/N, 
i.e., low-S/N spectral regions naturally contain fewer spikes. 
The effects of the spectral quality on our measurements are addressed 
by forward-modeled mock spectra. 

\input{tab2}

\section{Comparison to a hydrodynamical simulation}
\label{sec:trans_features_in_mock_spectra}
\subsection{Generation of Realistic Mock Spectra}
\label{sec:mock_spectra}

Following our procedure in \citetalias{Makan2021}, we use realistic 
forward-modeled mock spectra from a numerical simulation, applying 
different models for the \ion{He}{2} photoionization rate 
$\Gamma _{\mathrm{He\,II}}$ (i.e., a spatially constant UV background 
with varying amplitude, and a spatially fluctuating UV background). 
Only the comparison of the observational data to the mock spectra 
makes it possible to constrain the \ion{He}{2} reionization history 
using the transmission spikes and troughs. With this, we quantify 
the UV background in terms of $\Gamma _{\mathrm{He\,II}}$ 
similar to \citetalias{Davies2017} and \citetalias{Worseck2019}.

For the mock spectra we used skewers from \citetalias{Worseck2019}, 
which had been created from the outputs of a cubic (146\,cMpc)$^{3}$ 
hydrodynamical simulation run with \texttt{Nyx} code 
\citep{Almgren2013, Lukic2015} using the photoheating and 
photoionization rates from \citet{HaardtMadau2012}. For every skewer, 
the \ion{He}{2} Ly$\alpha$ optical depth was computed by using the 
velocity, temperature and rescaled density $\rho \propto (1+z)^{3}$ 
fields from the outputs of the simulation at redshifts 
$z_{\mathrm{sim}}$ = 2.2, 2.4, 2.5, 2.6, 3, 3.5, 4. 
\citetalias{Worseck2019} created 1000 $\Delta z = 0.08$ 
(60\,cMpc--100\,cMpc) long skewers in steps of $\mathrm{d}z = 0.04$ 
between $z = 2.28$ and $z = 3.88$.

In order to capture long troughs comparable to our 
observations, we constructed unique $2.28 \leqslant z \leqslant 3.88$ 
synthetic spectra by splicing random skewers. 
The skewers were 
spliced at the points of the same $\tau_\alpha(z)$ and similar 
derivatives $\mathrm{d}\tau_\alpha /\mathrm{d} z$ in the 
$\Delta z = 0.04$ long overlapping regions. In this way, we 
avoid discontinuities in $\tau_\alpha(z)$ between the skewers. 
Since observationally there must be no strong  
Lyman Limit systems along any of the 
\ion{He}{2}-transparent quasar sight lines, we excluded skewers 
with \ion{H}{1} Ly$\alpha$ optical depths $\tau_\mathrm{H\,I} > 3000$ 
which correspond to \ion{H}{1} column densities 
$N_{\mathrm{H\,I}} \gtrsim 10^{17.1}\mathrm{cm^{-2}}$. In this way, 
we created 1000 unique $2.28 \leqslant z \leqslant 3.88$ 
synthetic spectra with a pixel size of 2--3\,{$\mathrm{km\,s^{-1}}$} 
each containing 39 spliced skewers with an average length of $\Delta z\sim0.04$.

Assuming photoionization equilibrium in the optically thin 
limit, we rescaled the optical depths of the created spectra 
as $\tau_\alpha \propto \Gamma_\mathrm{HeII}^{-1}$
according to three sets of models:
(1) a spatially fluctuating redshift-dependent 
$\Gamma _{\mathrm{He\,II}}$ \citepalias{Davies2017}, 
(2) a spatially uniform $\Gamma _{\mathrm{He\,II}}$ that decreases 
with redshift as the median $\Gamma _{\mathrm{He\,II}}(z)$ from 
\citetalias{Davies2017} and 
(3) a set of spatially uniform $\Gamma _{\mathrm{He\,II}}$ with 
variable amplitude ($10^{-16}\mathrm{\,s^{-1}} 
\leqslant \Gamma _{\mathrm{He\,II}} \leqslant 
10^{-13}\mathrm{\,s^{-1}}$ with step size $\Delta[\log 
(\Gamma _{\mathrm{He\,II}}/\mathrm{s^{-1}})] = 0.1$).
The fluctuating UV background model from \citetalias{Davies2017} 
showed excellent agreement with the measured $\tau _{\mathrm{eff}}$ 
distribution at $z < 3.3$, although it is in modest tension with 
the data at $z>3.5$ \citepalias{Worseck2019}. The model was 
created using (500\,cMpc)$^{3}$ volume with an analytic IGM 
absorber model applying the \ion{H}{1} column density 
distribution model from \citet{Prochaska2014} and the quasar 
luminosity function from \citet{Hopkins2007}. The semi-analytical 
approach with approximate radiative transfer resulted in a 
spatially varying mean free path of ionizing photons, manifesting 
itself in the fluctuating UV background and consequently 
$\Gamma _{\mathrm{He\,II}}$. Here, we use a more recent simulation 
run from \citetalias{Davies2017} with a slightly higher resolution 
of (7.8 cMpc)$^3$ per grid cell in comparison to (10 cMpc)$^3$ 
used in \citetalias{Davies2017}. For comparison, we also used a 
simplified model of a uniform UV background that decreases with 
redshift according to the median $\Gamma _{\mathrm{He\,II}}(z)$ 
from \citetalias{Davies2017}. This model captures the general 
redshift evolution of $\Gamma _{\mathrm{He\,II}}$ which is expected 
due to the redshift evolution of the quasar luminosity function 
\citep[e.g.,][]{Hopkins2007,Kulkarni2019b, Puchwein2019}.

Similar to \citepalias{Makan2021}, we infer $\Gamma _{\mathrm{He\,II}}$ 
by comparing the incidence of spikes in the observed and mock data 
on large spatial scales of 150\,cMpc. Here we use the set of uniform 
UV background models \citepalias{Makan2021}. By doing so, we neglect 
spatial fluctuations in the UV background, which is justified because 
our 150\,cMpc scale is much larger than the mean free path of 
\ion{He}{2}-ionizing photons at these redshifts \citepalias{Davies2017}.
This set of uniform UV background models is used only in Sections 
\ref{sec:gamma_from_incidence} and \ref{sec:unif_uv_bkg}.

For all used UV background models the 1000 synthetic spectra were 
forward-modeled to resemble the observed spectra. For this, 
we applied the wavelength dependent HST/COS G130M line-spread 
function to the transmission $e^{-\tau_\alpha}$, rebinned it to 
the wavelength grid of the observed spectra and used the continuum, 
calibration curve, pixel exposure time and background to transform 
the transmission into expected counts per pixel. Measured counts 
were simulated by adding Poisson noise according to the expected 
counts per pixel. The resulting 1000 mock spectra per quasar sight 
line and model fully resemble the observational data in spectral 
resolution and quality including data gaps, enabling a one-to-one 
comparison.

\subsection{The \ion{He}{2} Trough Length Distribution}
\label{sec:troughs_in_mocks}

\begin{figure}
    \label{fig:troughs_2d_dist}
	\includegraphics[width = \linewidth]{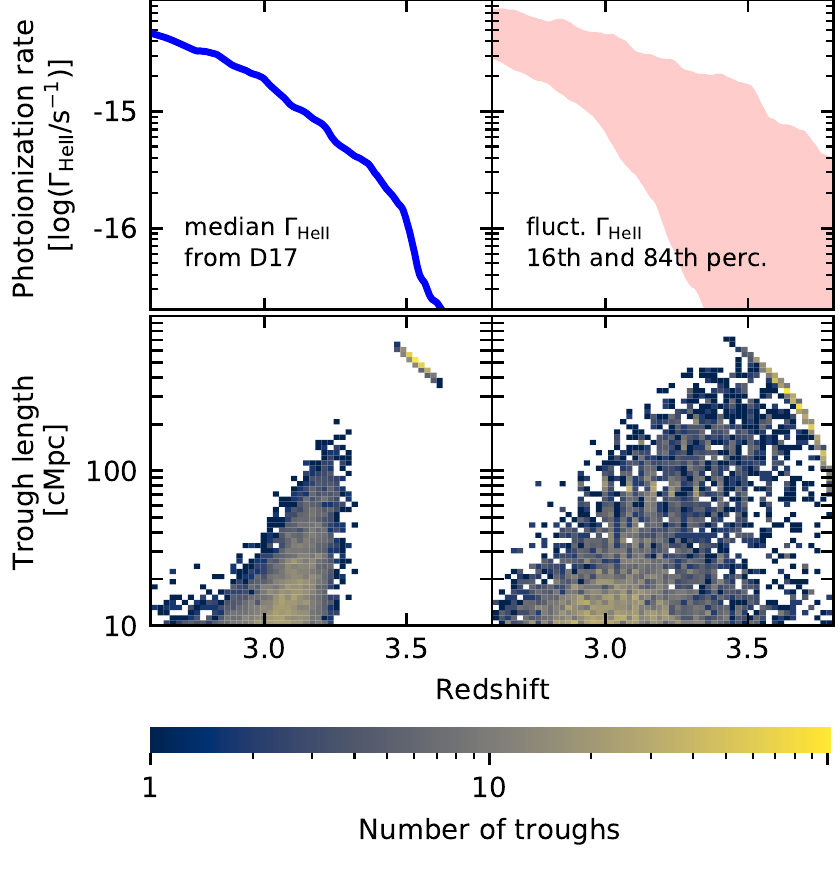}
	\caption{
	Redshift evolution of $\Gamma _{\mathrm{He\,II}}$ 
	(upper panels) and the predicted number of troughs and their 
	lengths (lower panels) in noise-free synthetic spectra for 
	the spatially uniform $\Gamma _{\mathrm{He\,II}}$ model (left) 
	and the fluctuating $\Gamma _{\mathrm{He\,II}}$ model (right).
	The stripe in the top right corner is caused by the redshift 
	boundary of our synthetic 
    spectra, naturally limiting the measured lengths of the troughs. 
	}
\end{figure}

For the fluctuating $\Gamma _{\mathrm{He\,II}}$ model and the 
median $\Gamma _{\mathrm{He\,II}}(z)$ from \citetalias{Davies2017} 
we measured the trough lengths in the 1000 noise-free synthetic 
spectra to determine their redshift evolution without any 
observational effects. 
This means that we model the HST/COS count distributions from 
the transmission by using simplified wavelength-independent 
sensitivity curves, exposure times and dark currents 
to calculate $P(>N | B)$, but we do not add Poisson noise to them. 
The results are presented in 
Figure~\ref{fig:troughs_2d_dist}. For the purpose of this plot, 
we defined the redshift position of the troughs as the midpoint in 
redshift $\Delta z$. 
The stripe in the upper right corner of both lower panels 
is due to the truncation of the troughs at the highest redshift of 
the synthetic spectra ($z=3.88$). 
At $z>3.3$, the uniform UV background model has such a low amplitude 
($\Gamma _{\mathrm{He\,II}} \lesssim 10^{-15.4}\mathrm{\,s^{-1}}$) 
that only very long troughs are produced, which disagrees with the observations. 
Due to the midpoint definition of the troughs this results 
in a gap at $z\sim3.4$. 
The fluctuating $\Gamma _{\mathrm{He\,II}}$ model provides a more 
realistic redshift evolution of the troughs with a wide range 
of trough lengths even at the highest redshifts.

\begin{figure}
	\includegraphics[width = \linewidth]{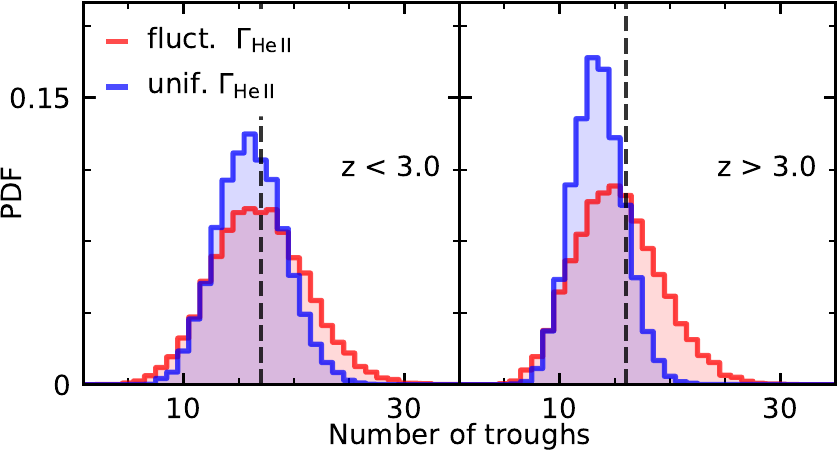}
	\caption{PDFs of the number of troughs with $L \geqslant 10$\,cMpc 
	in 10000 sample realizations at $z < 3$ and $z > 3$. 
	Results for the \citetalias{Davies2017} fluctuating UV background 
	model are shown in red, while those for a spatially uniform 
	median $\Gamma _{\mathrm{He\,II}}(z)$ from \citetalias{Davies2017} 
	are shown in blue. The numbers of detected troughs in the observed 
	sample are indicated with the dashed lines.}
	\label{fig:number_of_troughs_dist}
\end{figure}

Due to the small number of observed troughs (33 out 
of which 16 are at $z > 3$), it is not 
possible to capture their redshift evolution in detail. Instead, 
we investigated the distribution of troughs (number and lengths) 
separately at high ($z > 3$) and low ($z < 3$) redshift. 
Figure~\ref{fig:number_of_troughs_dist} shows the probability 
density functions (PDFs) of the number of troughs per mock sample 
(eight spectra) for the two $\Gamma _{\mathrm{He\,II}}$ models. 
For this, we used 10000 sample realizations of the mock data. 
Both models are consistent with the measured number of troughs 
given our data. Given the width and similarity of the predicted 
distributions, the number of troughs is insufficient to distinguish 
between the two models.

\begin{figure}
	\includegraphics[width = \linewidth]{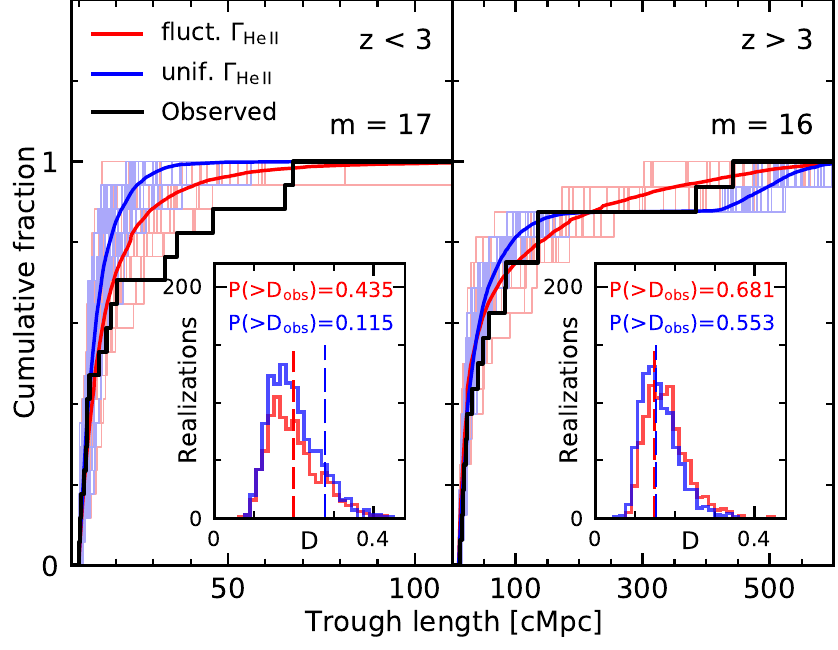}
	\caption{CDFs of the trough lengths in the observed sample (black) 
	and in the respective mock samples 
	(red: spatially fluctuating $\Gamma _{\mathrm{He\,II}}$; 
	blue: spatially uniform $\Gamma _{\mathrm{He\,II}}(z)$). 
	The solid red and blue lines were constructed from 10000 sample 
	realizations using all samples with $m=17$ troughs at 
	$z < 3.0$ (left panel) and $m=16$ at $z>3.0$ (right panel).
	The pale blue and red lines show 25 representative realizations 
	for the respective models. The inset shows the distributions of 
	the K-S test statistic $D$ for the CDFs of the individual 
	realizations and the CDF of all samples. The dashed lines mark 
	the values $D_\mathrm{obs}$ for the observed data and all samples. 
	$P(>D_\mathrm{obs}) > 0.1$ indicates that the observed data is 
	consistent with both models at 0.1 significance level.
	}
	\label{fig:troughs_cumul_dist}
\end{figure}

\begin{figure}
	\includegraphics[width = \linewidth]{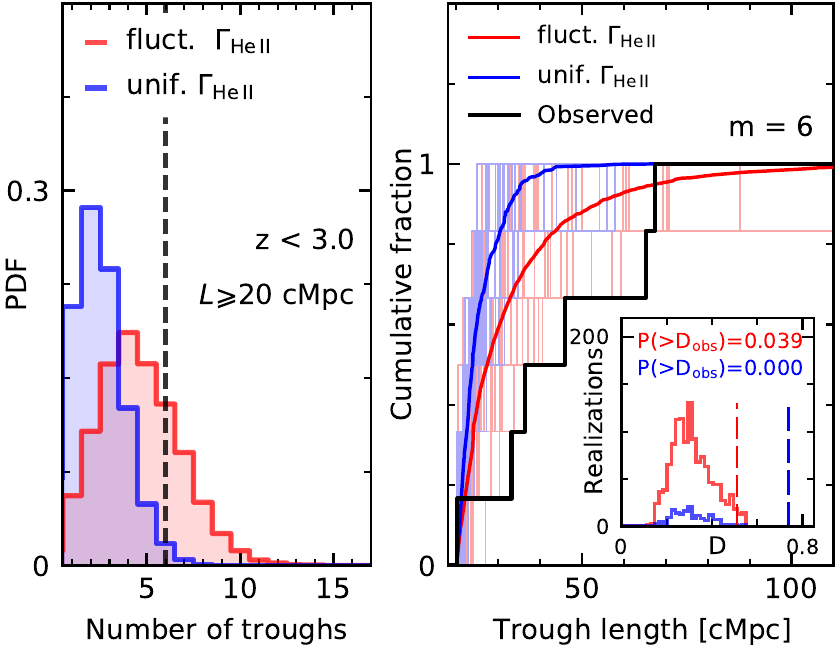}
	\caption{Similar to Figures \ref{fig:number_of_troughs_dist} and 
	\ref{fig:troughs_cumul_dist}, but for $ L \geqslant 20$\,cMpc at $z<3$. 
	}
	\label{fig:troughs_cumul_dist_20Mpc}
\end{figure}

\begin{figure}
    \label{fig:spikes_vs_gamma}
	\includegraphics[width = \linewidth]{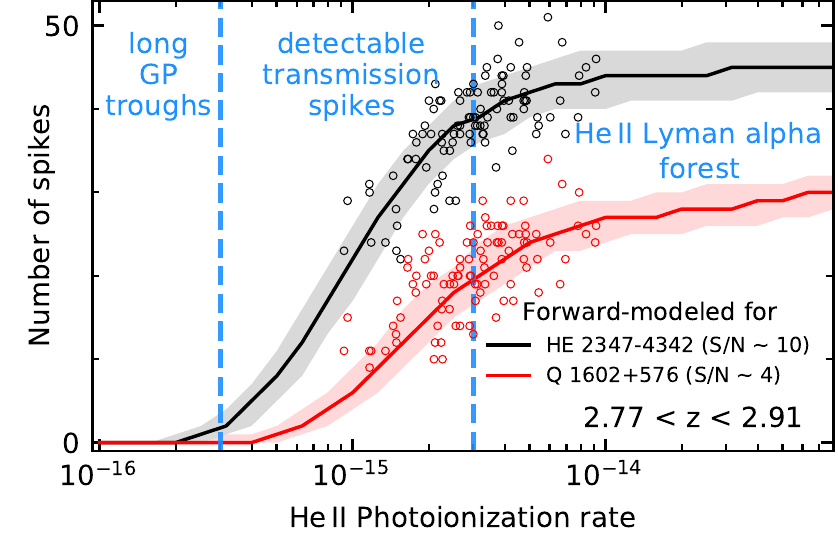}
	\caption{Increase in the number of \ion{He}{2} Ly$\alpha$ 
	transmission spikes with increasing photoionization rate 
	$\Gamma _{\mathrm{He\,II}}$ in 1000 mock spectra at $2.77 < z < 2.91$ 
	(150\,cMpc) at spectral quality of HE\,2347$-$4342 (black) and 
	Q\,1602$+$576 (red). The lines show the median number of spikes for 
	an assumed spatially uniform $\Gamma _{\mathrm{He\,II}}$ with their 
	respective 84th and 16th percentiles (shades). The circles show the 
	measured number of spikes in 100 representative mock spectra with 
	the fluctuating UV background model plotted at the respective median 
	$\Gamma _\mathrm{He\,II,\,D17}$ of the skewer. 
	Here, many more transmission spikes are measured toward 
	HE\,2347$-$4342 due to the higher S/N. 
	}
\end{figure}

\begin{figure*}\label{fig:posteriors_high_redshift}
	\includegraphics[width = \textwidth]{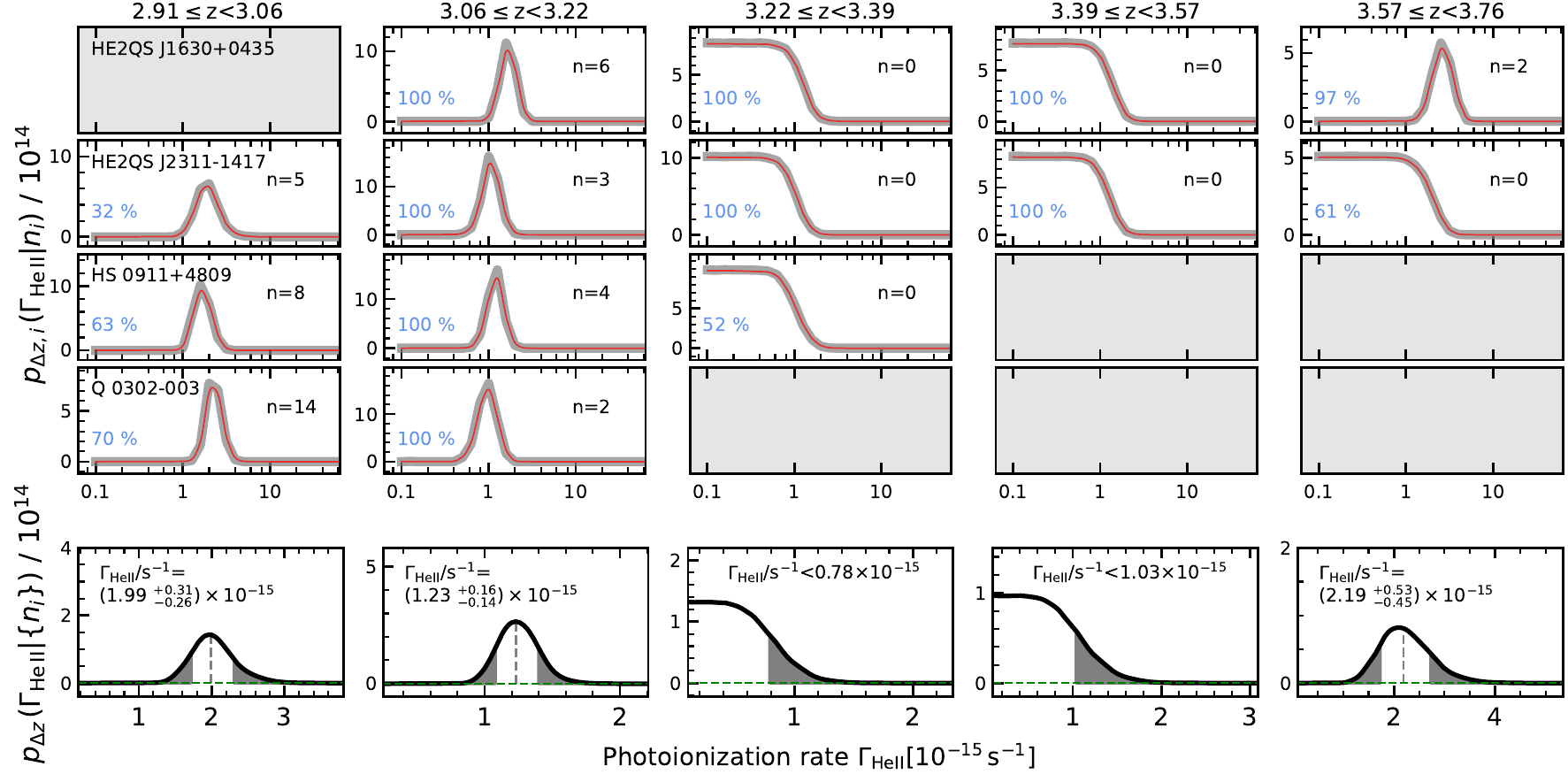}
	\caption{Top: Normalized posteriors 
	$p_{\Delta z, i} (\Gamma _{\mathrm{He\,II}} | n_i)$ for 150\,cMpc 
	long redshift bins at $z \geqslant 2.91$. The posteriors sampled 
	at $\Delta \log(\Gamma_{\mathrm{He\,II}}/\mathrm{s^{-1}}) = 0.1$ 
	are shown in gray. Final results were obtained after smoothing 
	and subsampling them with a Gaussian kernel 
	($\sigma = 0.05\times10^{-15}\mathrm{\,s^{-1}}$, red). 
	The grayed out panels indicate discarded redshift 
	bins (usable data in $<30\,\%$ of the 
	bin, i.e., $<45\,$cMpc). In all other panels we indicate the 
	fractional data coverage. Bottom: Joint posteriors 
	$p_{\Delta z} (\Gamma _{\mathrm{He\,II}} | \{n_i\})$. Vertical 
	dashed lines indicate the posterior median while the gray areas 
	mark the tails below the 16th and above the 84th percentile, 
	yielding the measurements and statistical uncertainties of
	$\Gamma _{\mathrm{He\,II}}$, respectively.}
\end{figure*}

While the number of troughs does not evolve with redshift, 
the troughs are becoming significantly longer with increasing 
redshift. Figure~\ref{fig:troughs_cumul_dist} shows the cumulative 
distribution functions (CDFs) of the trough lengths for the 
observed (Table~\ref{tab:troughs}) and the mock spectra. For the 
computation of the CDFs from mocks, we selected sample realizations 
with the same number of troughs as in the observed data ($m=17$ at 
$z<3$ and $m=16$ at $z>3$). This selection is required, because 
the distribution of trough lengths changes with the number of 
troughs whose lengths are constrained by the total available path 
length.

At $z<3$, the modeled CDFs do not seem to match the observed CDF,
due to the five $L>30$\,cMpc troughs. For the fluctuating 
$\Gamma _\mathrm{He\,II}$ model, only $\sim$3\% of our sample 
realizations have at least five $L>30$\,cMpc troughs, 
while there are almost none for the uniform $\Gamma _\mathrm{He\,II}$ 
model. At $z>3$, both models agree with the observed data. 
Given our sample size and quality, the modeled CDFs 
are too close to be distinguishable.

We performed a two-sample Kolmogorov-Smirnov (K-S) test to 
quantitatively compare the observed data to the models. First, we 
computed the K-S test statistic $D$ (absolute maximum distance 
between two CDFs) for each CDF from the individual sample 
realizations (pale colors in Figure~\ref{fig:troughs_cumul_dist}) 
and the CDF of all sample realizations (dark colors). The resulting 
distributions of $D$ for each model are shown in the insets of 
Figure~\ref{fig:troughs_cumul_dist}. Then, by calculating the 
fraction of $D > D_\mathrm{obs}$ where $D_\mathrm{obs}$ is the 
K-S test statistic between the observed data and the model, we 
estimated the probability $P(>D_\mathrm{obs})$ that the model 
is consistent with the observations. At any redshift both models 
cannot be rejected ($P(>D_\mathrm{obs})>0.1$). However, for the 
uniform UV background model at $z<3$, none of the mock samples 
features as many long troughs 
($30\mathrm{\,cMpc} \lesssim L \lesssim 70\mathrm{\,cMpc}$) 
as the observations. A closer investigation reveals that the 
calculated K-S test statistic is dominated by sample variance at 
$L \lesssim 20$\,cMpc. 
The PDFs and CDFs recalculated for $L\geqslant 20$\,cMpc 
do not change significantly at $z>3$. At $z<3$ however,  
the comparison of the PDFs in Figure \ref{fig:troughs_cumul_dist_20Mpc} 
clearly favors the fluctuating UV background model. 
Nevertheless, the CDFs of both models are not consistent 
with the observations owing to the small probability 
$P(>D_\mathrm{obs}) < 0.04$. 
The formal rejection of the \citetalias{Davies2017} fluctuating 
UV background model suggests that the UV background might fluctuate 
on larger scales (or stronger) than predicted by \citetalias{Davies2017}. 
Overall, short troughs with $L \lesssim 20$\,cMpc that occur 
frequently and may be impacted by density fluctuations hamper 
the model distinction, similar to \citet{Zhu2021} for \ion{H}{1}.

\subsection{Inference of the \ion{He}{2} Photoionization Rate}
\label{sec:inference_of_gamma}

\subsubsection{The Incidence of Spikes Depends on $\Gamma _\mathrm{He\,II}$}
\label{sec:gamma_from_incidence}

We showed in \citetalias{Makan2021} that the incidence of 
transmission spikes can be used to infer the volume-weighted 
median \ion{He}{2} photoionization rate. 
As a representative example, Figure~\ref{fig:spikes_vs_gamma} 
shows how the number of spikes increases with increasing 
$\Gamma _{\mathrm{He\,II}}$ for mock spectra of two quasars 
with different data quality in the redshift range $z = 2.77$--$2.91$ 
(150\,cMpc) with the spatially uniform (lines) and fluctuating 
(circles) UV background models. For the uniform 
$\Gamma _\mathrm{He\,II}$ model, the graph can be roughly divided 
in three regions. At $\Gamma _{\mathrm{He\,II}} 
\lesssim 3\times10^{-16}\mathrm{\,s^{-1}}$, there are only a few 
spikes, and the spectra are dominated by troughs. Then at 
$\Gamma _{\mathrm{He\,II}} \gtrsim 3\times10^{-16}\mathrm{\,s^{-1}}$, 
spikes become more frequent, first in the most under-dense regions. 
Here, the number of spikes increases monotonically. The curve 
reaches a plateau at $\Gamma _{\mathrm{He\,II}} 
\sim (3$--$6)\times10^{-15}\mathrm{\,s^{-1}}$, as the number of 
detectable spikes at a given data quality saturates. At higher 
$\Gamma _{\mathrm{He\,II}}$, the mock spectra become predominantly 
transmitting, showing an emerging \ion{He}{2} Ly$\alpha$ forest. 
The plot shows that the incidence of spikes is very sensitive to 
the \ion{He}{2} photoionization rate $\Gamma _{\mathrm{He\,II}}$ in 
the regime of detectable transmission spikes. The small variance 
comes mainly from IGM density fluctuations and Poisson noise in 
the mock spectra.

To show the dependence of the spike incidence on a 
fluctuating $\Gamma _\mathrm{He\,II}$ in Figure~\ref{fig:spikes_vs_gamma}, 
we used 100 representative mock spectra for the same sight lines 
and redshift bin. In Figure~\ref{fig:spikes_vs_gamma}, we plot 
the calculated median \ion{He}{2} photoionization rate per sight line
$\Gamma _\mathrm{He\,II,\,D17}^\mathrm{\,los}$ and the measured 
number of spikes for each of these mock spectra. The range of 
$\Gamma _\mathrm{He\,II,\,D17}^\mathrm{\,los}$ is predefined 
by the fixed model and thus, does not cover the full range of our 
custom uniform $\Gamma _\mathrm{He\,II}$ models. The number of 
spikes increases with increasing $\Gamma _\mathrm{He\,II,\,D17}^\mathrm{\,los}$ 
in a similar manner and with similar scatter as the uniform UV background 
models. The range of $\Gamma _\mathrm{He\,II,\,D17}^\mathrm{\,los}$ 
changes with redshift but always follows the curves for the 
spatially uniform UV background models. Therefore, approximating 
the fluctuating UV background with a set of uniform  models is 
well justified on our adopted large scale of 150\,cMpc.

\begin{figure}\label{fig:posteriors_low_redshift}
	\includegraphics[width = \linewidth]{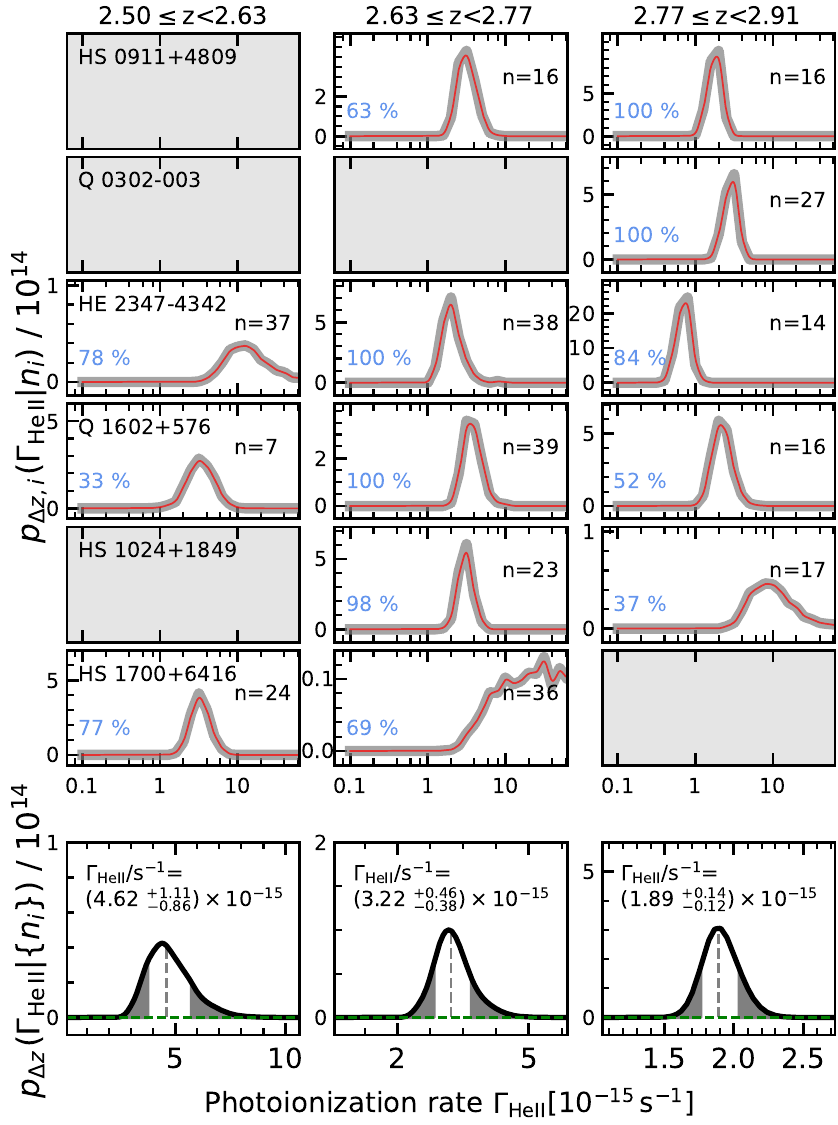}
	\caption{ Similar to Figure~\ref{fig:posteriors_high_redshift}, 
	but for redshift bins at $z \leqslant 2.91$.
	}
\end{figure}

\subsubsection{Inference of $\Gamma _{\mathrm{He\,II}}$ for the Uniform UV Background}
\label{sec:unif_uv_bkg}

Similar to \citetalias{Makan2021}, we used the number of 
transmission spikes in predefined redshift bins to infer the 
photoionization rate $\Gamma _{\mathrm{He\,II}}$. For this, we 
calculated the likelihood 
$\mathcal{L}(n | \Gamma _{\mathrm{He\,II}})$ from the PDFs of 
the number of spikes $n$ in fixed 150\,cMpc long redshift bins 
using our mock spectra for the set of uniform UV background 
models. Then by using Bayes' theorem, we constructed a 
normalized posterior 
\begin{equation}
    p_{\Delta z, i} (\Gamma _{\mathrm{He\,II}} | n_i) \propto 
    \mathcal{L}_{\Delta z, i} (n_i | \Gamma _{\mathrm{He\,II}})p(\Gamma _{\mathrm{He\,II}})
\end{equation}
given the number of detected spikes $n_i$ in the redshift bin 
$\Delta z$ along the sight line $i$ using a flat prior 
$p(\Gamma _{\mathrm{He\,II}})$ in the range of our UV 
background models ($10^{-16}\mathrm{\,s^{-1}} \leqslant 
\Gamma _{\mathrm{He\,II}} \leqslant 10^{-13}\mathrm{\,s^{-1}}$). 
The calculations were performed for every sight line and 
redshift bin that has usable data in $\geqslant 30\,\%$ of 
the bin ($\geqslant 45$\,cMpc). The used bins size of 150\,cMpc 
ensures a good sampling of transmission spikes (especially a 
concern at $z > 3$) while tracking the redshift evolution 
of the \ion{He}{2} photoionization rate. Their positions were 
chosen to maximize the usable spectral coverage, resulting 
in eight redshift bins at $ 2.5 \leqslant z \leqslant 3.76$.

The top panels of Figures \ref{fig:posteriors_high_redshift} 
and \ref{fig:posteriors_low_redshift} show the resulting 
posteriors. Since there are no spikes detected at  
$3.22 \leqslant z < 3.57$ along any of the sight lines, the 
posteriors provide only upper limits on 
$\Gamma _{\mathrm{He\,II}}$. Here we are in the regime of 
long troughs where the \ion{He}{2} photoionization rate at 
these redshifts must be $\Gamma _{\mathrm{He\,II}} \lesssim 
2\times 10^{-15}\,\mathrm{s^{-1}}$ given our data. A good 
example of an opposite extreme case is at $z=2.63$--$2.77$ 
toward HS\,1700+6416. Here, we are in the \ion{He}{2} Ly$\alpha$ 
forest regime (the plateaus at high $\Gamma _{\mathrm{He\,II}}$ 
values in Figure~\ref{fig:spikes_vs_gamma}) where the detected 
number of spikes is not sensitive to the photoionization rate 
anymore. Only two other posteriors show this behavior, at 
$z=2.50$--$2.63$ toward HE\,2347$-$4342 and at $z=2.77$--$2.91$ 
toward HS\,1024$+$1849.

Most of the posteriors are well defined, providing inferences 
of the photoionization rate. The positions of the posteriors 
at the same redshift sometimes varies substantially, exceeding 
expectations from IGM density fluctuations that dominate the 
widths of individual posteriors. This is further evidence 
for a fluctuating UV background, possibly down to $z\sim 2.6$.

We inferred the median photoionization rate $\Gamma _{\mathrm{He\,II}}$ 
in every redshift bin from the joint posterior 
\begin{equation}\label{eq:joint_posterior}
    p_{\Delta z} (\Gamma _{\mathrm{He\,II}} | \{n_i\}) \propto 
    \prod _{i} p_{\Delta z, i} (\Gamma _{\mathrm{He\,II}} | n_i)\quad.
\end{equation}
To ensure a good sampling of 
$p_{\Delta z} (\Gamma _{\mathrm{He\,II}} | \{n_i\})$ we 
smoothed and subsampled the individual posteriors with 
a $\sigma = 0.05\times10^{-15}\mathrm{\,s^{-1}}$ Gaussian kernel to reduce 
effects of partial overlap over the $\Gamma _{\mathrm{He\,II}}$ 
range. A single joint posterior distribution estimates the most 
probable \ion{He}{2} photoionization rate along all contributing 
sight lines for our assumed uniform UV background. 
We quote the median of $p_{\Delta z}$ 
(Figures \ref{fig:posteriors_high_redshift} and 
\ref{fig:posteriors_low_redshift}, bottom panel) as our measurement 
of $\Gamma _{\mathrm{He\,II}}$, while the 16th and 84th percentile 
yield the equal-tailed $1\sigma$ uncertainty. For the redshift bins 
without detected transmission spikes we quote the 84th percentile 
as the $1\sigma$ upper limit. Table~\ref{tab:results_gamma} lists 
the resulting photoionization rates $\Gamma _{\mathrm{He\,II}}$.

\subsubsection{Inference of $\Gamma _{\mathrm{He\,II}}$ for the Fluctuating UV Background}
\label{sec:gamma_for_fluctuating_uvb}

\begin{figure}
    \label{fig:delta_gamma}
	\includegraphics[width = \linewidth]{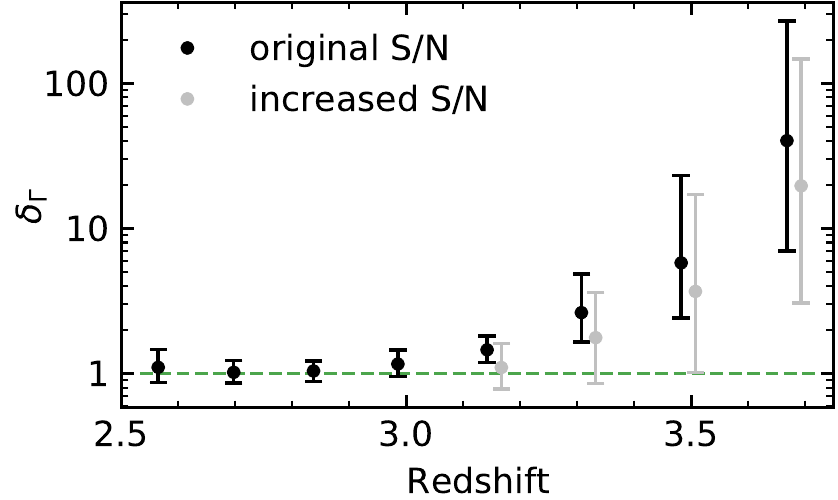}
	\caption{Median ratio between the measured photoionization 
	rate $\Gamma_{\mathrm{He\,II}}$ and the median 
	$\Gamma_{\mathrm{He\,II, D17}}$,
	$\delta _\Gamma = 
	\Gamma_{\mathrm{He\,II}}/\Gamma _\mathrm{He\,II,\,D17} 
	^{\mathrm{\,med}}$ (black), as determined from 1000 mock 
	sample realizations. Gray symbols show the case for an 
	increased S/N$\sim$15--30 at the highest redshifts.
	}
\end{figure}

Although the approximation of a uniform UV background along 
individual sight lines is well justified 
(see Section~\ref{sec:gamma_from_incidence}), it might be 
too strong for the inference of $\Gamma _{\mathrm{He\,II}}$ 
from the joint posterior. As a result, this joint posterior 
may lead to unrealistic constraints on $\Gamma_{\mathrm{He\,II}}$.
To test whether the joint posterior yields a representative 
value for the more realistic fluctuating UV background from 
\citetalias{Davies2017}, we inferred $\Gamma _{\mathrm{He\,II}}$ 
for each of the 1000 sample realizations in our defined redshift bins. 
Then we compared the inferred $\Gamma _{\mathrm{He\,II}}$ to the known 
median photoionization rate $\Gamma _\mathrm{He\,II,\,D17} 
^{\mathrm{\,med}}$ in the same sample realizations by 
calculating $\delta _\Gamma = 
\Gamma_{\mathrm{He\,II}}/\Gamma_{\mathrm{He\,II,\, D17}}
^{\mathrm{\,med}}$. We caution that 
$\Gamma_{\mathrm{He\,II,\, D17}}^{\mathrm{\,med}}$ can vary 
from sample to sample, i.e.\ it is not the overall median 
\ion{He}{2} photoionization rate used in 
Section~\ref{sec:troughs_in_mocks}.

Figure~\ref{fig:delta_gamma} shows the median $\delta _\Gamma$ 
with its 16th--84th percentile range. At $z<3$ the inferred
$\Gamma _{\mathrm{He\,II}}$ agrees with the input 
$\Gamma_{\mathrm{He\,II, D17} }^\mathrm{\,med} $ to within a 
small scatter, showing that Equation~\ref{eq:joint_posterior} 
is still applicable to a fluctuating UV background. 
At $z>3$, however, the \ion{He}{2} photoionization rate 
is systematically overestimated. A closer investigation shows 
that the bias appears if the joint likelihood is computed for 
spectra with and without spikes (i.e., at $3.57\leqslant z<3.76$ 
in Figure~\ref{fig:posteriors_high_redshift}). In these cases, 
the $n>0$ posterior will always narrow the joint posterior, 
whereby the lower $\Gamma _\mathrm{He\,II}$ values from the 
$n=0$ posteriors are effectively ignored. This artificially 
increases the inferred $\Gamma _\mathrm{He\,II}$, such that 
$\Gamma _\mathrm{He\,II}$ is representative only for ionized 
(i.e., transmissive) regions of the IGM. Due to the dropping 
\citetalias{Davies2017} \ion{He}{2} photoionization rate the 
bias strongly increases with redshift.
This bias can be somewhat 
mitigated by a higher S/N, such that transmission spikes are 
detectable in many spectra (Figure~\ref{fig:delta_gamma}). 
It completely disappears if all spectra show transmission spikes 
($3.06\leqslant z <3.22$ in Figure~\ref{fig:delta_gamma}).

The combination of $n>0$ and $n=0$ posteriors in or data appears 
only in the highest-redshift bin ($3.57 \leqslant z < 3.76$), 
making this measurement unreliable. 
Due to the well-defined posteriors at $z < 3.22$, our procedure 
yields the median $\Gamma_{\mathrm{He\,II}}$ also in the case of 
a fluctuating UV background. In Table~\ref{tab:results_gamma}, 
we list the inferred $\Gamma _{\mathrm{He\,II}}$ values, 
which correspond to the sample median $\Gamma_{\mathrm{He\,II}}$ 
of the contributing observed sight lines. 

\begin{deluxetable}{cccc}
\tablecaption{Inferred \ion{He}{2} 
Photoionization Rates ($\Gamma _{\mathrm{He\,II}}$) in 
Redshift Bins $\Delta z$ from $s$ Sight Lines with a Total Path Length $d$.}
\label{tab:results_gamma}
\tablehead{
\colhead{$\Delta z$} & 
\colhead{$\Gamma _{\mathrm{He\,II}}(10^{-15}\mathrm{s^{-1}})$} &
\colhead{$d$ (cMpc)} & 
\colhead{$s$}}
\tabletypesize{\footnotesize}
\startdata	
2.50--2.63  &   $4.62^{+1.11}_{-0.86}$      &   280 &   3   \\ 
2.63--2.77  &   $3.22^{+0.46}_{-0.38}$      &   655 &   5   \\ 
2.77--2.91  &   $1.89^{+0.14}_{-0.12}$      &   559 &   5   \\ 
2.91--3.06  &   $1.99^{+0.31}_{-0.26}$      &   258 &   3   \\ 
3.06--3.22  &   $1.23^{+0.16}_{-0.14}$      &   600 &   4   \\ 
3.22--3.39  &   $ \leqslant 0.78$           &   379 &   3   \\ 
3.39--3.57  &   $ \leqslant 1.03$           &   300 &   2   \\ 
3.57--3.76  &   $2.19^{+0.53}_{-0.45}$ $^\ast$&   237 &   2   \\ 
\enddata
\tablenotetext{^\ast}{Unreliable.}
\end{deluxetable}

\begin{figure}
    \label{fig:gamma_redshift_evolution}
	\includegraphics[width = \linewidth]{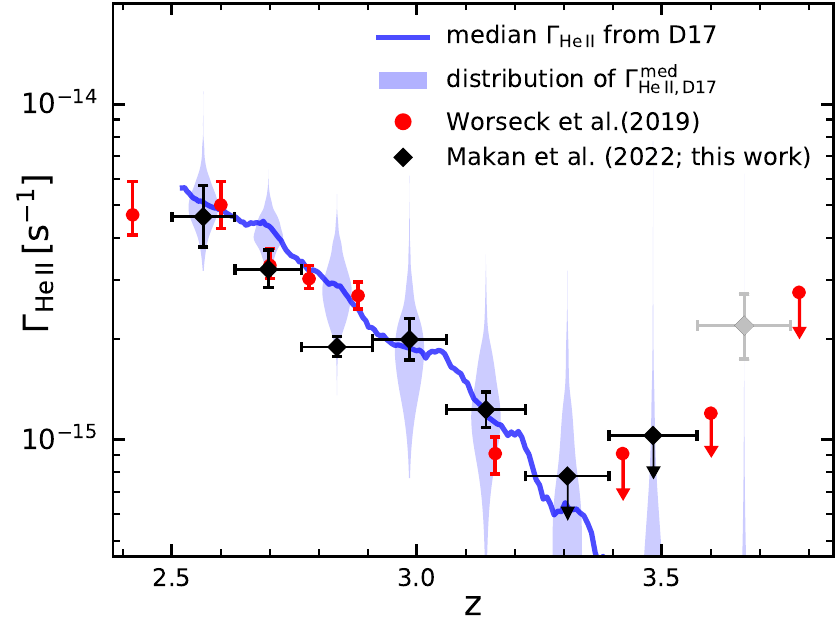} 
	\caption{Redshift evolution of the \ion{He}{2} photoionization 
	rate $\Gamma _{\mathrm{He\,II}}$ inferred from the incidence of 
	transmission spikes (black) and from the \ion{He}{2} effective 
	optical depth \citepalias[red, ][]{Worseck2019}.
	The unreliable inference at $z\sim3.7$ is showed in gray. 
	The line indicates the median $\Gamma _\mathrm{He\,II,\,D17}$ 
	while the violins show the distribution of the sample medians
	$\Gamma _\mathrm{He\,II,\,D17} ^{\mathrm{\,med}}$ 
	from 1000 sample realizations.}
\end{figure}

\subsection{Redshift Evolution of the \ion{He}{2} Photoionization Rate}

Figure~\ref{fig:gamma_redshift_evolution} shows the 
redshift evolution of the inferred \ion{He}{2} photoionization 
rates in comparison to previously published measurements by 
\citetalias{Worseck2019} and to the fluctuating UV background 
model by \citetalias{Davies2017}.
The \ion{He}{2} photoionization rate strongly decreases 
from $z \simeq 2.5$ to $z \simeq 3.2$,
and likely even at higher redshifts,
 as indicated by our upper limits on $\Gamma _{\mathrm{He\,II}}$.
As previously mentioned, the measurement at $z \sim 3.7$ is considered 
untrustworthy due to the strong bias uncertainties and thus, it will 
not be discussed further.

For the comparison to the predicted $\Gamma _{\mathrm{He\,II}}$ from the 
\citetalias{Davies2017} model, we considered the strong variance from our 
limited sample size. For this, we computed the sample medians 
$\Gamma_{\mathrm{He\,II,\, D17}}^{\mathrm{\,med}}$ from corresponding 
subsamples of the 1000 \citetalias{Davies2017} $\Gamma _{\mathrm{He\,II}}$ 
skewers (Section~\ref{sec:mock_spectra}). The violins in 
Figure~\ref{fig:gamma_redshift_evolution} show the distribution of 
1000 calculated $\Gamma_{\mathrm{He\,II,\, D17}}^{\mathrm{\,med}}$ values, 
indicating that substantial scatter from sample variance is expected. 
Our inferences are in very good agreement with the \citetalias{Davies2017} 
model given the expected sample variance, except at $z\simeq 2.8$. 
The reason for the outlier at $z\simeq 2.8$ is the unusual 
low-transmission spectrum of HE\,2347$-$4342 (Figures~\ref{fig:low_z_spectra} 
and \ref{fig:posteriors_low_redshift}). In particular, the 36\,cMpc long 
trough at $z\sim2.83$ is difficult to reproduce in \ion{He}{2} reionization 
models, possibly indicating a very rare region with a still significant 
\ion{He}{2} fraction \citep[e.g.,][]{Shull2010, FurlanettoDixon2010}. 
The exclusion of this sight line from our small sample results in 
$\Gamma _{\mathrm{He\,II}} = (2.53 ^{+0.34} _{-0.27}) 
\times 10^{-15}\mathrm{\,s^{-1}}$ at $z\simeq 2.8$ 
making it consistent with \citetalias{Davies2017}.

Figure~\ref{fig:gamma_redshift_evolution} also shows the 
$\Gamma _{\mathrm{He\,II}}$ inferences from the effective optical 
depth measurements by \citetalias{Worseck2019}. Our measurements are 
in broad agreement with \citetalias{Worseck2019}, with the largest 
difference at $z \simeq 2.8$ due to the particular sight line of 
HE\,2347$-$4342. Additionally, unlike \citetalias{Worseck2019}, we are 
able to measure $\Gamma _{\mathrm{He\,II}}$ at $z\sim 3$ owing to 
the longer spectral coverage and the negligible scattered light 
around geocoronal Ly$\alpha$ in our G130M data.

\section{Summary}
\label{sec:summary}

We have systematically analyzed small-scale and large-scale \ion{He}{2} 
Ly$\alpha$ absorption features (transmission spikes and absorption 
troughs) in eight high-resolution ($R \simeq 12,500$--$18,000$) 
HST/COS spectra. This is the first comprehensive sample of 
high-resolution \ion{He}{2} absorption spectra probing the end of 
the \ion{He}{2} reionization epoch at $ 2.50 < z < 3.76$. We measured 
trough lengths and the incidence of spikes using the automated 
routine established in \citepalias{Makan2021}. Then, we compared our 
measurements to predictions from forward-modeled mock spectra from a 
hydrodynamical simulation employing different UV background models.

We measured 33 troughs (i.e., $L\geqslant 10$\,cMpc regions without 
significant small-scale transmission), 16 of which are at $z>3$.
While the number of troughs is similar at $z<3$ and $z>3$, 
their length dramatically increases with increasing redshift. 
There are six $82\mathrm{\,cMpc} \leqslant L \leqslant 442$\,cMpc 
troughs at $z>3$, whereas each trough at $z<3$ spans $\lesssim 65$\,cMpc. 
The distribution of trough lengths at $z < 3$ is in disagreement with 
the predictions from a spatially uniform UV background and the 
\citepalias{Davies2017} fluctuating UV background model, suggesting that 
the UV background might fluctuate on even larger scales than predicted 
by \citetalias{Davies2017}. At $z>3$, our measured trough length 
distribution cannot rule out the tested models, although the UV 
background is naturally expected to fluctuate strongly at these 
redshifts (Figure~\ref{fig:number_of_troughs_dist}).

We used the incidence of transmission spikes to infer \ion{He}{2} 
photoionization rate $\Gamma _{\mathrm{He\,II}}$ similar to 
\citetalias{Makan2021}. For this we compared the measured 
incidence to predictions from a (146\,cMpc)$^3$ hydrodynamical 
simulation for a wide range of $\Gamma _{\mathrm{He\,II}}$. The inferred 
$\Gamma _{\mathrm{He\,II}}$ decreases with increasing redshift from 
$\simeq 4.6\times 10^{-15}\mathrm{\,s^{-1}}$ at $z\simeq 2.6$ to 
$\simeq 1.2 \times 10^{-15}\mathrm{\,s^{-1}}$ at $z\simeq3.2$, in 
agreement with the \citetalias{Davies2017} fluctuating UV background 
model (Figure~\ref{fig:gamma_redshift_evolution}). At $3.22 < z < 3.57$, 
the lack of transmission spikes results in upper limits 
$\Gamma _{\mathrm{He\,II}} \lesssim 10^{-15}\mathrm{\,s^{-1}}$ that 
are still in agreement with \citetalias{Davies2017}. At these high 
redshifts we find that our measurement technique overestimates the 
IGM \ion{He}{2} photoionization rate, because it is dominated by 
transmission spikes that probe rare highly ionized regions. We have 
identified such a case at the highest redshifts probed here 
($3.57 \geqslant z < 3.76$). 
Conversely, due to the limited sensitivity of the 
\ion{He}{2} Ly$\alpha$ absorption to high \ion{He}{2} fractions, 
the troughs may be produced by large incompletely ionized regions similar to the suggested \ion{H}{1} ``neutral islands'' at $z\sim5.5$ 
\citep{Malloy2015, Kulkarni2019, Keating2020a}. 
Detailed comparisons to our data require refined numerical simulations 
with a broad range of \ion{He}{2} reionization histories.

We conclude that the incidence of transmission spikes is sensitive 
to the \ion{He}{2} photoionization rate (Figure~\ref{fig:spikes_vs_gamma}) 
and yields similar values as the \ion{He}{2} effective optical depth 
from larger but lower-resolution samples \citepalias{Worseck2019}.
On the other hand, we have shown that the trough statistics require a 
much larger sample to distinguish between different plausible models 
of \ion{He}{2} reionization and/or the fluctuating \ion{He}{2}-ionizing 
background. This is similar to the situation of \ion{H}{1} at 
$z \simeq 6$ where the trough statistics had been used early on 
\citep{SongailaCowie2002, Fan2006, Gallerani2006}, but only recently 
they had the ability to distinguish between reionization models in 
larger samples \citep{Zhu2021}. Given the substantial HST/COS exposure 
times and the generally faint background quasars, a significant 
increase of the sample requires a next-generation large UV space 
telescope \citep[e.g.,][]{LUVOIR2019}. A 6--8\,m UV-sensitive space 
telescope, as recently prioritized in the Decadal Survey on Astronomy 
and Astrophysics 2020 \citep{NAP2021}, would be able to gather 
high-resolution $R \simeq 20,000$ spectra of more than 20 
$m _\mathrm{FUV} = 21$--22 quasars discovered with HST 
\citep{Syphers2012,Worseck2019}. Such observations would statistically 
determine the evolution of the \ion{He}{2} absorption from individual 
high-redshift transmission spikes to a resolved \ion{He}{2} Ly$\alpha$ forest.

\acknowledgments
We thank the anonymous referee for the constructive review. 
This work was funded by Bundesministerium f\"ur Wirtschaft und Energie in 
the framework of the Verbundforschung of the Deutsches Zentrum f\"ur Luft- 
und Raumfahrt (DLR, grant 50 OR 1813). 
Support for program GO~15356 was provided by NASA through a grant from the 
Space Telescope Science Institute, which is operated by the Association of 
Universities for Research in Astronomy, Inc., under NASA contract NAS5-26555.
This research is based on observations made with the NASA/ESA Hubble Space 
Telescope obtained from the Space Telescope Science Institute, which is 
operated by the Association of Universities for Research in Astronomy, Inc., 
under NASA contract NAS 5–26555. These observations are associated with 
programs GTO~11528, GTO~12033, GO~12816, GO~13301 and GO~15356.

\facility{\textit{HST} (COS)}

\software{astropy \citep{Astropy2013,Astropy2018},
          SciPy \citep{Scipy2020},
          matplotlib \citep{Hunter2007},
          numpy \citep{Harris2020}
          }

 \clearpage
 \bibliography{references}

\end{document}

%% file: tab1.tex
\begin{deluxetable*}{lllccclcclclr}
\tablecaption{Sample of the \ion{He}{2}-transparent Quasars Observed with HST/COS G130M.}
\label{tab:quasar_sample}
\tablewidth{0pt}
\setlength{\tabcolsep}{1ex}
\renewcommand{\arraystretch}{1.0}
\tablehead{
\colhead{Object} & 
\colhead{R.A. (J2000)} & 
\colhead{Decl. (J2000)} & 
\colhead{$z_{\mathrm{em}}$\tablenotemark{a}}  & 
\colhead{$t_{\mathrm{exp}}$(s)} &  
\colhead{S/N\tablenotemark{b}} &
\colhead{$R$\tablenotemark{c}} &
\colhead{$\Delta \lambda$(\AA)} &
\colhead{PID\tablenotemark{d}} & 
\colhead{$f_{1500\mathrm{\,\AA}}$\tablenotemark{e}} & 
\colhead{$\alpha$\tablenotemark{f}}  & 
\colhead{$z_{\mathrm{abs}}$} & 
\colhead{$\log N_{\mathrm{H\,I}}$\tablenotemark{g}}}
\tabletypesize{\footnotesize}
\startdata	
HS\,1700$+$6416		& $17^{\mathrm{h}}01^{\mathrm{m}}00^{\mathrm{s}}.61$ & $+64\degr 12'09''.1$	& 2.7472    &	59,547	&  20 & 12500 & 1060--1370 & 13301 & 1.8836   & -2.027   & 0.8642 & 16.09\\
 & & & & & & & & & & & 0.7217 & 16.23\\
 & & & & & & & & & & & 0.5523 & 15.77\\
HS\,1024$+$1849		& $10^{\mathrm{h}}27^{\mathrm{m}}34^{\mathrm{s}}.13$ & $+18\degr 34'27''.6$	& 2.8521    &	28,689	&  5 & 12500 & 1060--1370 & 12816 & 3.224   & -0.732    & ... & ...\\
Q1602$+$576			& $16^{\mathrm{h}}03^{\mathrm{m}}55^{\mathrm{s}}.93$ & $+57\degr 30'54''.4$ & 2.8608    &	15,613	& 7  & 12500 & 1060--1370 & 12816 & 6.019   & -2.701    & ... & ...\\
HE\,2347$-$4342		& $23^{\mathrm{h}}50^{\mathrm{m}}34^{\mathrm{s}}.23$ & $-43\degr 25'59''.8$	& 2.8852    &	28,455	& 28 & 14500 & 1060--1444 & 11528 & 17.424   & -2.520    & ... & ...\\
					& &		&	 &	41,392	  & & & &  13301 & &  & ... & ...\\
Q0302$-$003			& $03^{\mathrm{h}}04^{\mathrm{m}}49^{\mathrm{s}}.86$ & $-00\degr 08'13''.5$	& 3.2850    &	24,869  & 5  & 18000 & 1125--1465 & 12033 & 2.879   & -3.534    & ... & ...\\
HS\,0911$+$4809		& $09^{\mathrm{h}}15^{\mathrm{m}}10^{\mathrm{s}}.01$ & $+47\degr 56'58''.8$	& 3.350     &	26,863  & 6  & 12500 & 1060--1370 & 12816 & 3.163   & -0.572    & ... & ...\\
HE2QS\,J2311$-$1417	& $23^{\mathrm{h}}11^{\mathrm{m}}45^{\mathrm{s}}.46$ & $-14\degr 17'52''.2$ & 3.700\tablenotemark{$\ast$}    &	54,920	& 3 & 14000 & 1080--1473 & 15356 & 2.065   & -2.600  & 0.7515 & 16.37\\
  & & & & & & & && & & 0.4779 & $<17.2$\\
HE2QS\,J1630$+$0435	& $16^{\mathrm{h}}30^{\mathrm{m}}56^{\mathrm{s}}.33$ & $+04\degr 35'59''.4$ & 3.810\tablenotemark{$\ast$}    &	45,915	& 3 & 14000 & 1080--1473 & 15356 & 2.955   & -1.158  & ... & ...\\
\enddata
\tablenotetext{a}{Redshift reference: \citet{Worseck2021}. The two sightlines indicated with $^\ast$  are from \citet{Khrykin2019}.}
\tablenotetext{b}{Poisson S/N per pixel of the co-added spectrum in the continuum near the \ion{He}{2} Ly$\alpha$ in the quasar rest frame.}
\tablenotetext{c}{Resolving power $R=\Delta \lambda / \lambda$ at 1250\,\AA.}
\tablenotetext{d}{{\it HST} program number.}
\tablenotetext{e}{Continuum flux density at 1500\,\AA\, in $10^{-16}\mathrm{\,erg\,cm^{-2}\,s^{-1}\,\AA^{-1}}$.}
\tablenotetext{f}{Power-law spectral index $\alpha$ for $E_{\mathrm{\lambda}} = f_{1500\mathrm{\,\AA}}(\lambda / 1500\mathrm{\,\AA})^{\alpha}$}
\tablenotetext{g}{Logarithmic column density of identified intervening \ion{H}{1} absorber at $z_{\mathrm{abs}}$ in cm$^{-2}$}
\end{deluxetable*}

%% file: tab2.tex
\begin{deluxetable}{lccl}
\tablecaption{Measured Troughs with the Length $L$ 
and Effective Optical Depth $\tau _ {\mathrm{eff}}$}
\label{tab:troughs}
\tablewidth{0pt}
\setlength{\tabcolsep}{1ex}
\renewcommand{\arraystretch}{1.0}
\tablehead{
\colhead{Sightline} & 
\colhead{$\Delta z$}  & 
\colhead{$L$ [cMpc]} &  
\colhead{$\tau _ {\mathrm{eff}}$}}
\startdata	
HS\,1700$+$6416     &   ...    &   ...  & ...    \\
HS\,1024$+$1849     &   2.691 - 2.722   &   33  & $3.87^{+0.46}_{-0.31}$ \\
                    &   2.723 - 2.734   &   12  & $2.99^{+0.27}_{-0.21}$ \\
                    &   2.741 - 2.757   &   17  & $>4.84$ \\
Q1602$+$576         &   2.590 - 2.600   &   12  & $>3.52$ \\
                    &   2.610 - 2.619   &   10  & $>3.97$ \\
                    &   2.798 - 2.810   &   13  & $>5.88$ \\
HE\,2347$-$4342     &   2.752 - 2.767   &   15  & $>7.63$ \\
                    &   2.769 - 2.788   &   20  & $>8.02$ \\
                    &   2.789 - 2.799   &   10  & $5.96^{+0.20}_{-0.16}$ \\
                    &   2.825 - 2.860   &   33  & $>8.56$ \\
Q0302$-$003         &   2.779 - 2.789   &   11  & $2.91^{+0.27}_{-0.21}$ \\
                    &   2.861 - 2.925   &   65  & $4.11^{+0.22}_{-0.18}$ \\
                    &   2.926 - 2.945   &   19  & $4.71^{+1.02}_{-0.50}$ \\
                    &   3.031 - 3.048   &   17  & $3.30^{+0.17}_{-0.14}$ \\
                    &   3.056 - 3.200   &   135 & $5.63^{+0.58}_{-0.37}$ \\
                    &   3.202 - 3.222   &   18  & $>5.05$    \\
HS\,0911$+$4809     &   2.673 - 2.684   &   12  & $3.70^{+1.13}_{-0.52}$ \\
                    &   2.749 - 2.792   &   46  & $3.64^{+0.20}_{-0.17}$ \\
                    &   2.793 - 2.804   &   12  & $3.96^{+0.57}_{-0.36}$ \\
                    &   2.867 - 2.934   &   67  & $6.05^{+4.26}_{-0.66}$ \\
                    &   3.029 - 3.052   &   22  & $3.78^{+0.23}_{-0.18}$ \\
                    &   3.054 - 3.067   &   13  & $3.82^{+0.28}_{-0.22}$ \\
                    &   3.069 - 3.159   &   84  & $4.90^{+0.32}_{-0.24}$ \\
                    &   3.163 - 3.179   &   15  & $4.76^{+0.67}_{-0.39}$ \\
                    &   3.188 - 3.279   &   82  & $>6.31$ \\
                    &   3.302 - 3.329   &   23  & $5.22^{+1.30}_{-0.57}$ \\
HE2QS\,J2311$-$1417 &   3.026 - 3.086   &   58  & $5.49^{+0.67}_{-0.39}$ \\
                    &   3.088 - 3.131   &   40  & $4.78^{+0.33}_{-0.25}$ \\
                    &   3.132 - 3.166   &   31  & $5.30^{+0.75}_{-0.42}$ \\
                    &   3.167 - 3.688   &  442  & $>6.83$ \\
HE2QS\,J1630$+$0435 &   3.084 - 3.135   &   48  & $5.37^{+1.28}_{-0.54}$ \\
                    &   3.138 - 3.580   &   383  & $>6.57$ \\
                    &   3.584 - 3.757   &   135  & $5.76^{+2.23}_{-0.65}$ \\
\enddata
\end{deluxetable}

%% file: km_final.bbl
\begin{thebibliography}{}
\expandafter\ifx\csname natexlab\endcsname\relax\def\natexlab#1{#1}\fi
\providecommand{\url}[1]{\href{#1}{#1}}
\providecommand{\dodoi}[1]{doi:~\href{http://doi.org/#1}{\nolinkurl{#1}}}
\providecommand{\doeprint}[1]{\href{http://ascl.net/#1}{\nolinkurl{http://ascl.net/#1}}}
\providecommand{\doarXiv}[1]{\href{https://arxiv.org/abs/#1}{\nolinkurl{https://arxiv.org/abs/#1}}}

\bibitem[{{Almgren} {et~al.}(2013){Almgren}, {Bell}, {Lijewski}, {Luki{\'c}},
  \& {Van Andel}}]{Almgren2013}
{Almgren}, A.~S., {Bell}, J.~B., {Lijewski}, M.~J., {Luki{\'c}}, Z., \& {Van
  Andel}, E. 2013, \apj, 765, 39, \dodoi{10.1088/0004-637X/765/1/39}

\bibitem[{{Astropy Collaboration} {et~al.}(2013){Astropy Collaboration},
  {Robitaille}, {Tollerud}, {Greenfield}, {Droettboom}, {Bray}, {Aldcroft},
  {Davis}, {Ginsburg}, {Price-Whelan}, {Kerzendorf}, {Conley}, {Crighton},
  {Barbary}, {Muna}, {Ferguson}, {Grollier}, {Parikh}, {Nair}, {Unther},
  {Deil}, {Woillez}, {Conseil}, {Kramer}, {Turner}, {Singer}, {Fox}, {Weaver},
  {Zabalza}, {Edwards}, {Azalee Bostroem}, {Burke}, {Casey}, {Crawford},
  {Dencheva}, {Ely}, {Jenness}, {Labrie}, {Lim}, {Pierfederici}, {Pontzen},
  {Ptak}, {Refsdal}, {Servillat}, \& {Streicher}}]{Astropy2013}
{Astropy Collaboration}, {Robitaille}, T.~P., {Tollerud}, E.~J., {et~al.} 2013,
  \aap, 558, A33, \dodoi{10.1051/0004-6361/201322068}

\bibitem[{{Astropy Collaboration} {et~al.}(2018){Astropy Collaboration},
  {Price-Whelan}, {Sip{\H{o}}cz}, {G{\"u}nther}, {Lim}, {Crawford}, {Conseil},
  {Shupe}, {Craig}, {Dencheva}, {Ginsburg}, {VanderPlas}, {Bradley},
  {P{\'e}rez-Su{\'a}rez}, {de Val-Borro}, {Aldcroft}, {Cruz}, {Robitaille},
  {Tollerud}, {Ardelean}, {Babej}, {Bach}, {Bachetti}, {Bakanov}, {Bamford},
  {Barentsen}, {Barmby}, {Baumbach}, {Berry}, {Biscani}, {Boquien}, {Bostroem},
  {Bouma}, {Brammer}, {Bray}, {Breytenbach}, {Buddelmeijer}, {Burke},
  {Calderone}, {Cano Rodr{\'\i}guez}, {Cara}, {Cardoso}, {Cheedella}, {Copin},
  {Corrales}, {Crichton}, {D'Avella}, {Deil}, {Depagne}, {Dietrich}, {Donath},
  {Droettboom}, {Earl}, {Erben}, {Fabbro}, {Ferreira}, {Finethy}, {Fox},
  {Garrison}, {Gibbons}, {Goldstein}, {Gommers}, {Greco}, {Greenfield},
  {Groener}, {Grollier}, {Hagen}, {Hirst}, {Homeier}, {Horton}, {Hosseinzadeh},
  {Hu}, {Hunkeler}, {Ivezi{\'c}}, {Jain}, {Jenness}, {Kanarek}, {Kendrew},
  {Kern}, {Kerzendorf}, {Khvalko}, {King}, {Kirkby}, {Kulkarni}, {Kumar},
  {Lee}, {Lenz}, {Littlefair}, {Ma}, {Macleod}, {Mastropietro}, {McCully},
  {Montagnac}, {Morris}, {Mueller}, {Mumford}, {Muna}, {Murphy}, {Nelson},
  {Nguyen}, {Ninan}, {N{\"o}the}, {Ogaz}, {Oh}, {Parejko}, {Parley}, {Pascual},
  {Patil}, {Patil}, {Plunkett}, {Prochaska}, {Rastogi}, {Reddy Janga},
  {Sabater}, {Sakurikar}, {Seifert}, {Sherbert}, {Sherwood-Taylor}, {Shih},
  {Sick}, {Silbiger}, {Singanamalla}, {Singer}, {Sladen}, {Sooley},
  {Sornarajah}, {Streicher}, {Teuben}, {Thomas}, {Tremblay}, {Turner},
  {Terr{\'o}n}, {van Kerkwijk}, {de la Vega}, {Watkins}, {Weaver}, {Whitmore},
  {Woillez}, {Zabalza}, \& {Astropy Contributors}}]{Astropy2018}
{Astropy Collaboration}, {Price-Whelan}, A.~M., {Sip{\H{o}}cz}, B.~M., {et~al.}
  2018, \aj, 156, 123, \dodoi{10.3847/1538-3881/aabc4f}

\bibitem[{{Barnett} {et~al.}(2017){Barnett}, {Warren}, {Becker}, {Mortlock},
  {Hewett}, {McMahon}, {Simpson}, \& {Venemans}}]{Barnett2017}
{Barnett}, R., {Warren}, S.~J., {Becker}, G.~D., {et~al.} 2017, \aap, 601, A16,
  \dodoi{10.1051/0004-6361/201630258}

\bibitem[{{Becker} {et~al.}(2015){Becker}, {Bolton}, {Madau}, {Pettini},
  {Ryan-Weber}, \& {Venemans}}]{Becker2015}
{Becker}, G.~D., {Bolton}, J.~S., {Madau}, P., {et~al.} 2015, \mnras, 447,
  3402, \dodoi{10.1093/mnras/stu2646}

\bibitem[{{Boera} {et~al.}(2014){Boera}, {Murphy}, {Becker}, \&
  {Bolton}}]{Boera2014}
{Boera}, E., {Murphy}, M.~T., {Becker}, G.~D., \& {Bolton}, J.~S. 2014, \mnras,
  441, 1916, \dodoi{10.1093/mnras/stu660}

\bibitem[{{Bosman} {et~al.}(2021){Bosman}, {Davies}, {Becker}, {Keating},
  {Davies}, {Zhu}, {Eilers}, {D'Odorico}, {Bian}, {Bischetti}, {Cristiani},
  {Fan}, {Farina}, {Haehnelt}, {Kulkarni}, {Mesinger}, {Meyer}, {Onoue},
  {Pallottini}, {Qin}, {Ryan-Weber}, {Schindler}, {Walter}, {Wang}, \&
  {Yang}}]{Bosman2021}
{Bosman}, S. E.~I., {Davies}, F.~B., {Becker}, G.~D., {et~al.} 2021, arXiv
  e-prints, arXiv:2108.03699.
\newblock \doarXiv{2108.03699}

\bibitem[{{Cardelli} {et~al.}(1989){Cardelli}, {Clayton}, \&
  {Mathis}}]{Cardelli1989}
{Cardelli}, J.~A., {Clayton}, G.~C., \& {Mathis}, J.~S. 1989, \apj, 345, 245,
  \dodoi{10.1086/167900}

\bibitem[{{Chardin} {et~al.}(2018){Chardin}, {Haehnelt}, {Bosman}, \&
  {Puchwein}}]{Chardin2018}
{Chardin}, J., {Haehnelt}, M.~G., {Bosman}, S. E.~I., \& {Puchwein}, E. 2018,
  \mnras, 473, 765, \dodoi{10.1093/mnras/stx2362}

\bibitem[{{Choudhury} {et~al.}(2021){Choudhury}, {Paranjape}, \&
  {Bosman}}]{Choudhury2021}
{Choudhury}, T.~R., {Paranjape}, A., \& {Bosman}, S. E.~I. 2021, \mnras, 501,
  5782, \dodoi{10.1093/mnras/stab045}

\bibitem[{{Compostella} {et~al.}(2013){Compostella}, {Cantalupo}, \&
  {Porciani}}]{Compostella2013}
{Compostella}, M., {Cantalupo}, S., \& {Porciani}, C. 2013, \mnras, 435, 3169,
  \dodoi{10.1093/mnras/stt1510}

\bibitem[{{Compostella} {et~al.}(2014){Compostella}, {Cantalupo}, \&
  {Porciani}}]{Compostella2014}
---. 2014, \mnras, 445, 4186, \dodoi{10.1093/mnras/stu2035}

\bibitem[{{Davies} {et~al.}(2017){Davies}, {Furlanetto}, \&
  {Dixon}}]{Davies2017}
{Davies}, F.~B., {Furlanetto}, S.~R., \& {Dixon}, K.~L. 2017, \mnras, 465,
  2886, \dodoi{10.1093/mnras/stw2868}

\bibitem[{{Eide} {et~al.}(2020){Eide}, {Ciardi}, {Graziani}, {Busch}, {Feng},
  \& {Di Matteo}}]{Eide2020}
{Eide}, M.~B., {Ciardi}, B., {Graziani}, L., {et~al.} 2020, \mnras, 498, 6083,
  \dodoi{10.1093/mnras/staa2774}

\bibitem[{{Eilers} {et~al.}(2019){Eilers}, {Hennawi}, {Davies}, \&
  {O{\~n}orbe}}]{Eilers2019}
{Eilers}, A.-C., {Hennawi}, J.~F., {Davies}, F.~B., \& {O{\~n}orbe}, J. 2019,
  \apj, 881, 23, \dodoi{10.3847/1538-4357/ab2b3f}

\bibitem[{{Fan} {et~al.}(2006){Fan}, {Strauss}, {Becker}, {White}, {Gunn},
  {Knapp}, {Richards}, {Schneider}, {Brinkmann}, \& {Fukugita}}]{Fan2006}
{Fan}, X., {Strauss}, M.~A., {Becker}, R.~H., {et~al.} 2006, \aj, 132, 117,
  \dodoi{10.1086/504836}

\bibitem[{{Fardal} {et~al.}(1998){Fardal}, {Giroux}, \& {Shull}}]{Fardal1998}
{Fardal}, M.~A., {Giroux}, M.~L., \& {Shull}, J.~M. 1998, \aj, 115, 2206,
  \dodoi{10.1086/300359}

\bibitem[{{Faucher-Gigu{\`e}re} {et~al.}(2008){Faucher-Gigu{\`e}re}, {Lidz},
  {Hernquist}, \& {Zaldarriaga}}]{FaucherGiguere2008}
{Faucher-Gigu{\`e}re}, C.-A., {Lidz}, A., {Hernquist}, L., \& {Zaldarriaga}, M.
  2008, \apj, 688, 85, \dodoi{10.1086/592289}

\bibitem[{{Fechner} \& {Reimers}(2007)}]{FechnerReimers2007}
{Fechner}, C., \& {Reimers}, D. 2007, \aap, 461, 847,
  \dodoi{10.1051/0004-6361:20065556}

\bibitem[{{Feldman} \& {Cousins}(1998)}]{FeldmanCousins1998}
{Feldman}, G.~J., \& {Cousins}, R.~D. 1998, \prd, 57, 3873,
  \dodoi{10.1103/PhysRevD.57.3873}

\bibitem[{{Furlanetto} \& {Dixon}(2010)}]{FurlanettoDixon2010}
{Furlanetto}, S.~R., \& {Dixon}, K.~L. 2010, \apj, 714, 355,
  \dodoi{10.1088/0004-637X/714/1/355}

\bibitem[{{Furlanetto} \& {Oh}(2008)}]{FurlanettoOh2008}
{Furlanetto}, S.~R., \& {Oh}, S.~P. 2008, \apj, 681, 1, \dodoi{10.1086/588546}

\bibitem[{{Gaikwad} {et~al.}(2020){Gaikwad}, {Rauch}, {Haehnelt}, {Puchwein},
  {Bolton}, {Keating}, {Kulkarni}, {Ir{\v{s}}i{\v{c}}}, {Ba{\~n}ados},
  {Becker}, {Boera}, {Zahedy}, {Chen}, {Carswell}, {Chardin}, \&
  {Rorai}}]{Gaikwad2020}
{Gaikwad}, P., {Rauch}, M., {Haehnelt}, M.~G., {et~al.} 2020, \mnras, 494,
  5091, \dodoi{10.1093/mnras/staa907}

\bibitem[{{Gallerani} {et~al.}(2006){Gallerani}, {Choudhury}, \&
  {Ferrara}}]{Gallerani2006}
{Gallerani}, S., {Choudhury}, T.~R., \& {Ferrara}, A. 2006, \mnras, 370, 1401,
  \dodoi{10.1111/j.1365-2966.2006.10553.x}

\bibitem[{{Gallerani} {et~al.}(2008){Gallerani}, {Ferrara}, {Fan}, \&
  {Choudhury}}]{Gallerani2008}
{Gallerani}, S., {Ferrara}, A., {Fan}, X., \& {Choudhury}, T.~R. 2008, \mnras,
  386, 359, \dodoi{10.1111/j.1365-2966.2008.13029.x}

\bibitem[{{Garaldi} {et~al.}(2019){Garaldi}, {Gnedin}, \&
  {Madau}}]{Garaldi2019}
{Garaldi}, E., {Gnedin}, N.~Y., \& {Madau}, P. 2019, \apj, 876, 31,
  \dodoi{10.3847/1538-4357/ab12dc}

\bibitem[{{Gnedin} {et~al.}(2017){Gnedin}, {Becker}, \& {Fan}}]{Gnedin2017}
{Gnedin}, N.~Y., {Becker}, G.~D., \& {Fan}, X. 2017, \apj, 841, 26,
  \dodoi{10.3847/1538-4357/aa6c24}

\bibitem[{{Green} {et~al.}(2012){Green}, {Froning}, {Osterman}, {Ebbets},
  {Heap}, {Leitherer}, {Linsky}, {Savage}, {Sembach}, {Shull}, {Siegmund},
  {Snow}, {Spencer}, {Stern}, {Stocke}, {Welsh}, {B{\'e}land}, {Burgh},
  {Danforth}, {France}, {Keeney}, {McPhate}, {Penton}, {Andrews},
  {Brownsberger}, {Morse}, \& {Wilkinson}}]{Green2012}
{Green}, J.~C., {Froning}, C.~S., {Osterman}, S., {et~al.} 2012, \apj, 744, 60,
  \dodoi{10.1088/0004-637X/744/1/60}

\bibitem[{{Haardt} \& {Madau}(2012)}]{HaardtMadau2012}
{Haardt}, F., \& {Madau}, P. 2012, \apj, 746, 125,
  \dodoi{10.1088/0004-637X/746/2/125}

\bibitem[{Harris {et~al.}(2020)Harris, Millman, van~der Walt, Gommers,
  Virtanen, Cournapeau, Wieser, Taylor, Berg, Smith, Kern, Picus, Hoyer, van
  Kerkwijk, Brett, Haldane, del R{\'{i}}o, Wiebe, Peterson,
  G{\'{e}}rard-Marchant, Sheppard, Reddy, Weckesser, Abbasi, Gohlke, \&
  Oliphant}]{Harris2020}
Harris, C.~R., Millman, K.~J., van~der Walt, S.~J., {et~al.} 2020, Nature, 585,
  357, \dodoi{10.1038/s41586-020-2649-2}

\bibitem[{{Hiss} {et~al.}(2018){Hiss}, {Walther}, {Hennawi}, {O{\~n}orbe},
  {O'Meara}, {Rorai}, \& {Luki{\'c}}}]{Hiss2018}
{Hiss}, H., {Walther}, M., {Hennawi}, J.~F., {et~al.} 2018, \apj, 865, 42,
  \dodoi{10.3847/1538-4357/aada86}

\bibitem[{{Hopkins} {et~al.}(2007){Hopkins}, {Richards}, \&
  {Hernquist}}]{Hopkins2007}
{Hopkins}, P.~F., {Richards}, G.~T., \& {Hernquist}, L. 2007, \apj, 654, 731,
  \dodoi{10.1086/509629}

\bibitem[{{Hu} {et~al.}(2019){Hu}, {Wang}, {Zheng}, {Malhotra}, {Rhoads},
  {Infante}, {Barrientos}, {Yang}, {Jiang}, {Kang}, {Perez}, {Wold}, {Hibon},
  {Jiang}, {Khostovan}, {Valdes}, {Walker}, {Galaz}, {Coughlin}, {Harish},
  {Kong}, {Pharo}, \& {Zheng}}]{Hu2019}
{Hu}, W., {Wang}, J., {Zheng}, Z.-Y., {et~al.} 2019, \apj, 886, 90,
  \dodoi{10.3847/1538-4357/ab4cf4}

\bibitem[{Hunter(2007)}]{Hunter2007}
Hunter, J.~D. 2007, Computing in Science \& Engineering, 9, 90,
  \dodoi{10.1109/MCSE.2007.55}

\bibitem[{{Jakobsen} {et~al.}(1994){Jakobsen}, {Boksenberg}, {Deharveng},
  {Greenfield}, {Jedrzejewski}, \& {Paresce}}]{Jakobsen1994}
{Jakobsen}, P., {Boksenberg}, A., {Deharveng}, J.~M., {et~al.} 1994, \nat, 370,
  35, \dodoi{10.1038/370035a0}

\bibitem[{{Jakobsen} {et~al.}(2003){Jakobsen}, {Jansen}, {Wagner}, \&
  {Reimers}}]{Jakobsen2003}
{Jakobsen}, P., {Jansen}, R.~A., {Wagner}, S., \& {Reimers}, D. 2003, \aap,
  397, 891, \dodoi{10.1051/0004-6361:20021579}

\bibitem[{{Keating} {et~al.}(2020){Keating}, {Weinberger}, {Kulkarni},
  {Haehnelt}, {Chardin}, \& {Aubert}}]{Keating2020a}
{Keating}, L.~C., {Weinberger}, L.~H., {Kulkarni}, G., {et~al.} 2020, \mnras,
  491, 1736, \dodoi{10.1093/mnras/stz3083}

\bibitem[{{Khrykin} {et~al.}(2016){Khrykin}, {Hennawi}, {McQuinn}, \&
  {Worseck}}]{Khrykin2016}
{Khrykin}, I.~S., {Hennawi}, J.~F., {McQuinn}, M., \& {Worseck}, G. 2016, \apj,
  824, 133, \dodoi{10.3847/0004-637X/824/2/133}

\bibitem[{{Khrykin} {et~al.}(2019){Khrykin}, {Hennawi}, \&
  {Worseck}}]{Khrykin2019}
{Khrykin}, I.~S., {Hennawi}, J.~F., \& {Worseck}, G. 2019, \mnras, 484, 3897,
  \dodoi{10.1093/mnras/stz135}

\bibitem[{{Kim} {et~al.}(2021){Kim}, {Wakker}, {Nasir}, {Carswell}, {Savage},
  {Bolton}, {Fox}, {Viel}, {Haehnelt}, {Charlton}, \& {Rosenwasser}}]{Kim2021}
{Kim}, T.~S., {Wakker}, B.~P., {Nasir}, F., {et~al.} 2021, \mnras, 501, 5811,
  \dodoi{10.1093/mnras/staa3844}

\bibitem[{{Kriss} {et~al.}(2001){Kriss}, {Shull}, {Oegerle}, {Zheng},
  {Davidsen}, {Songaila}, {Tumlinson}, {Cowie}, {Deharveng}, {Friedman},
  {Giroux}, {Green}, {Hutchings}, {Jenkins}, {Kruk}, {Moos}, {Morton},
  {Sembach}, \& {Tripp}}]{Kriss2001}
{Kriss}, G.~A., {Shull}, J.~M., {Oegerle}, W., {et~al.} 2001, Science, 293,
  1112, \dodoi{10.1126/science.1062693}

\bibitem[{{Kulkarni} {et~al.}(2019{\natexlab{a}}){Kulkarni}, {Keating},
  {Haehnelt}, {Bosman}, {Puchwein}, {Chardin}, \& {Aubert}}]{Kulkarni2019}
{Kulkarni}, G., {Keating}, L.~C., {Haehnelt}, M.~G., {et~al.}
  2019{\natexlab{a}}, \mnras, 485, L24, \dodoi{10.1093/mnrasl/slz025}

\bibitem[{{Kulkarni} {et~al.}(2019{\natexlab{b}}){Kulkarni}, {Worseck}, \&
  {Hennawi}}]{Kulkarni2019b}
{Kulkarni}, G., {Worseck}, G., \& {Hennawi}, J.~F. 2019{\natexlab{b}}, \mnras,
  488, 1035, \dodoi{10.1093/mnras/stz1493}

\bibitem[{{La Plante} {et~al.}(2017){La Plante}, {Trac}, {Croft}, \&
  {Cen}}]{LaPlante2017}
{La Plante}, P., {Trac}, H., {Croft}, R., \& {Cen}, R. 2017, \apj, 841, 87,
  \dodoi{10.3847/1538-4357/aa7136}

\bibitem[{{Luki{\'c}} {et~al.}(2015){Luki{\'c}}, {Stark}, {Nugent}, {White},
  {Meiksin}, \& {Almgren}}]{Lukic2015}
{Luki{\'c}}, Z., {Stark}, C.~W., {Nugent}, P., {et~al.} 2015, \mnras, 446,
  3697, \dodoi{10.1093/mnras/stu2377}

\bibitem[{{Madau} {et~al.}(1999){Madau}, {Haardt}, \& {Rees}}]{Madau1999}
{Madau}, P., {Haardt}, F., \& {Rees}, M.~J. 1999, \apj, 514, 648,
  \dodoi{10.1086/306975}

\bibitem[{{Madau} \& {Meiksin}(1994)}]{MadauMeiksin1994}
{Madau}, P., \& {Meiksin}, A. 1994, \apjl, 433, L53, \dodoi{10.1086/187546}

\bibitem[{{Makan} {et~al.}(2021){Makan}, {Worseck}, {Davies}, {Hennawi},
  {Prochaska}, \& {Richter}}]{Makan2021}
{Makan}, K., {Worseck}, G., {Davies}, F.~B., {et~al.} 2021, \apj, 912, 38,
  \dodoi{10.3847/1538-4357/abee17}

\bibitem[{{Malloy} \& {Lidz}(2015)}]{Malloy2015}
{Malloy}, M., \& {Lidz}, A. 2015, \apj, 799, 179,
  \dodoi{10.1088/0004-637X/799/2/179}

\bibitem[{{Mason} {et~al.}(2018){Mason}, {Treu}, {Dijkstra}, {Mesinger},
  {Trenti}, {Pentericci}, {de Barros}, \& {Vanzella}}]{Mason2018}
{Mason}, C.~A., {Treu}, T., {Dijkstra}, M., {et~al.} 2018, \apj, 856, 2,
  \dodoi{10.3847/1538-4357/aab0a7}

\bibitem[{{McQuinn}(2009)}]{McQuinn2009}
{McQuinn}, M. 2009, \apjl, 704, L89, \dodoi{10.1088/0004-637X/704/2/L89}

\bibitem[{{McQuinn} {et~al.}(2009){McQuinn}, {Lidz}, {Zaldarriaga},
  {Hernquist}, {Hopkins}, {Dutta}, \& {Faucher-Gigu{\`e}re}}]{McQuinn2009b}
{McQuinn}, M., {Lidz}, A., {Zaldarriaga}, M., {et~al.} 2009, \apj, 694, 842,
  \dodoi{10.1088/0004-637X/694/2/842}

\bibitem[{{McQuinn} \& {Worseck}(2014)}]{McQuinnWorseck2014}
{McQuinn}, M., \& {Worseck}, G. 2014, \mnras, 440, 2406,
  \dodoi{10.1093/mnras/stu242}

\bibitem[{{Meiksin}(2020)}]{Meiksin2020}
{Meiksin}, A. 2020, \mnras, 491, 4884, \dodoi{10.1093/mnras/stz3395}

\bibitem[{{Miralda-Escude}(1993)}]{Miralda-Escude1993}
{Miralda-Escude}, J. 1993, \mnras, 262, 273, \dodoi{10.1093/mnras/262.1.273}

\bibitem[{{Miralda-Escud{\'e}} {et~al.}(2000){Miralda-Escud{\'e}}, {Haehnelt},
  \& {Rees}}]{Miralda-Escude2000}
{Miralda-Escud{\'e}}, J., {Haehnelt}, M., \& {Rees}, M.~J. 2000, \apj, 530, 1,
  \dodoi{10.1086/308330}

\bibitem[{{M{\o}ller} \& {Jakobsen}(1990)}]{MollerJakobsen1990}
{M{\o}ller}, P., \& {Jakobsen}, P. 1990, \aap, 228, 299

\bibitem[{{Morrissey} {et~al.}(2007){Morrissey}, {Conrow}, {Barlow}, {Small},
  {Seibert}, {Wyder}, {Budav{\'a}ri}, {Arnouts}, {Friedman}, {Forster},
  {Martin}, {Neff}, {Schiminovich}, {Bianchi}, {Donas}, {Heckman}, {Lee},
  {Madore}, {Milliard}, {Rich}, {Szalay}, {Welsh}, \& {Yi}}]{Morrissey2007}
{Morrissey}, P., {Conrow}, T., {Barlow}, T.~A., {et~al.} 2007, \apjs, 173, 682,
  \dodoi{10.1086/520512}

\bibitem[{{Murthy}(2014)}]{Murthy2014}
{Murthy}, J. 2014, \apjs, 213, 32, \dodoi{10.1088/0067-0049/213/2/32}

\bibitem[{{National Academies of Sciences, Engineering, and
  Medicine}(2021)}]{NAP2021}
{National Academies of Sciences, Engineering, and Medicine}. 2021, Pathways to
  Discovery in Astronomy and Astrophysics for the 2020s (Washington, DC: The
  National Academies Press), \dodoi{10.17226/26141}.
\newblock
  \url{https://www.nap.edu/catalog/26141/pathways-to-discovery-in-astronomy-and-astrophysics-for-the-2020s}

\bibitem[{{Paschos} \& {Norman}(2005)}]{Paschos2005}
{Paschos}, P., \& {Norman}, M.~L. 2005, \apj, 631, 59, \dodoi{10.1086/431787}

\bibitem[{{Picard} \& {Jakobsen}(1993)}]{PicardJakobsen1993}
{Picard}, A., \& {Jakobsen}, P. 1993, \aap, 276, 331

\bibitem[{{Planck Collaboration} {et~al.}(2020){Planck Collaboration},
  {Aghanim}, {Akrami}, {Ashdown}, {Aumont}, {Baccigalupi}, {Ballardini},
  {Banday}, {Barreiro}, {Bartolo}, {Basak}, {Battye}, {Benabed}, {Bernard},
  {Bersanelli}, {Bielewicz}, {Bock}, {Bond}, {Borrill}, {Bouchet}, {Boulanger},
  {Bucher}, {Burigana}, {Butler}, {Calabrese}, {Cardoso}, {Carron},
  {Challinor}, {Chiang}, {Chluba}, {Colombo}, {Combet}, {Contreras}, {Crill},
  {Cuttaia}, {de Bernardis}, {de Zotti}, {Delabrouille}, {Delouis}, {Di
  Valentino}, {Diego}, {Dor{\'e}}, {Douspis}, {Ducout}, {Dupac}, {Dusini},
  {Efstathiou}, {Elsner}, {En{\ss}lin}, {Eriksen}, {Fantaye}, {Farhang},
  {Fergusson}, {Fernandez-Cobos}, {Finelli}, {Forastieri}, {Frailis},
  {Fraisse}, {Franceschi}, {Frolov}, {Galeotta}, {Galli}, {Ganga},
  {G{\'e}nova-Santos}, {Gerbino}, {Ghosh}, {Gonz{\'a}lez-Nuevo}, {G{\'o}rski},
  {Gratton}, {Gruppuso}, {Gudmundsson}, {Hamann}, {Handley}, {Hansen},
  {Herranz}, {Hildebrandt}, {Hivon}, {Huang}, {Jaffe}, {Jones}, {Karakci},
  {Keih{\"a}nen}, {Keskitalo}, {Kiiveri}, {Kim}, {Kisner}, {Knox},
  {Krachmalnicoff}, {Kunz}, {Kurki-Suonio}, {Lagache}, {Lamarre}, {Lasenby},
  {Lattanzi}, {Lawrence}, {Le Jeune}, {Lemos}, {Lesgourgues}, {Levrier},
  {Lewis}, {Liguori}, {Lilje}, {Lilley}, {Lindholm}, {L{\'o}pez-Caniego},
  {Lubin}, {Ma}, {Mac{\'\i}as-P{\'e}rez}, {Maggio}, {Maino}, {Mandolesi},
  {Mangilli}, {Marcos-Caballero}, {Maris}, {Martin}, {Martinelli},
  {Mart{\'\i}nez-Gonz{\'a}lez}, {Matarrese}, {Mauri}, {McEwen}, {Meinhold},
  {Melchiorri}, {Mennella}, {Migliaccio}, {Millea}, {Mitra},
  {Miville-Desch{\^e}nes}, {Molinari}, {Montier}, {Morgante}, {Moss}, {Natoli},
  {N{\o}rgaard-Nielsen}, {Pagano}, {Paoletti}, {Partridge}, {Patanchon},
  {Peiris}, {Perrotta}, {Pettorino}, {Piacentini}, {Polastri}, {Polenta},
  {Puget}, {Rachen}, {Reinecke}, {Remazeilles}, {Renzi}, {Rocha}, {Rosset},
  {Roudier}, {Rubi{\~n}o-Mart{\'\i}n}, {Ruiz-Granados}, {Salvati}, {Sandri},
  {Savelainen}, {Scott}, {Shellard}, {Sirignano}, {Sirri}, {Spencer},
  {Sunyaev}, {Suur-Uski}, {Tauber}, {Tavagnacco}, {Tenti}, {Toffolatti},
  {Tomasi}, {Trombetti}, {Valenziano}, {Valiviita}, {Van Tent}, {Vibert},
  {Vielva}, {Villa}, {Vittorio}, {Wandelt}, {Wehus}, {White}, {White},
  {Zacchei}, \& {Zonca}}]{PlanckCollab2018}
{Planck Collaboration}, {Aghanim}, N., {Akrami}, Y., {et~al.} 2020, \aap, 641,
  A6, \dodoi{10.1051/0004-6361/201833910}

\bibitem[{{Prochaska} {et~al.}(2014){Prochaska}, {Madau}, {O'Meara}, \&
  {Fumagalli}}]{Prochaska2014}
{Prochaska}, J.~X., {Madau}, P., {O'Meara}, J.~M., \& {Fumagalli}, M. 2014,
  \mnras, 438, 476, \dodoi{10.1093/mnras/stt2218}

\bibitem[{{Puchwein} {et~al.}(2019){Puchwein}, {Haardt}, {Haehnelt}, \&
  {Madau}}]{Puchwein2019}
{Puchwein}, E., {Haardt}, F., {Haehnelt}, M.~G., \& {Madau}, P. 2019, \mnras,
  485, 47, \dodoi{10.1093/mnras/stz222}

\bibitem[{{Reimers} {et~al.}(1997){Reimers}, {Kohler}, {Wisotzki}, {Groote},
  {Rodriguez-Pascual}, \& {Wamsteker}}]{Reimers1997}
{Reimers}, D., {Kohler}, S., {Wisotzki}, L., {et~al.} 1997, \aap, 327, 890.
\newblock \doarXiv{astro-ph/9707173}

\bibitem[{{Schlegel} {et~al.}(1998){Schlegel}, {Finkbeiner}, \&
  {Davis}}]{Schlegel1998}
{Schlegel}, D.~J., {Finkbeiner}, D.~P., \& {Davis}, M. 1998, \apj, 500, 525,
  \dodoi{10.1086/305772}

\bibitem[{{Schmidt} {et~al.}(2018){Schmidt}, {Hennawi}, {Worseck}, {Davies},
  {Luki{\'c}}, \& {O{\~n}orbe}}]{Schmidt2018}
{Schmidt}, T.~M., {Hennawi}, J.~F., {Worseck}, G., {et~al.} 2018, \apj, 861,
  122, \dodoi{10.3847/1538-4357/aac8e4}

\bibitem[{{Schmidt} {et~al.}(2017){Schmidt}, {Worseck}, {Hennawi}, {Prochaska},
  \& {Crighton}}]{Schmidt2017}
{Schmidt}, T.~M., {Worseck}, G., {Hennawi}, J.~F., {Prochaska}, J.~X., \&
  {Crighton}, N. H.~M. 2017, \apj, 847, 81, \dodoi{10.3847/1538-4357/aa83ac}

\bibitem[{{Shull} \& {Danforth}(2020)}]{ShullDanforth2020}
{Shull}, J.~M., \& {Danforth}, C.~W. 2020, \apj, 899, 163,
  \dodoi{10.3847/1538-4357/aba3c9}

\bibitem[{{Shull} {et~al.}(2010){Shull}, {France}, {Danforth}, {Smith}, \&
  {Tumlinson}}]{Shull2010}
{Shull}, J.~M., {France}, K., {Danforth}, C.~W., {Smith}, B., \& {Tumlinson},
  J. 2010, \apj, 722, 1312, \dodoi{10.1088/0004-637X/722/2/1312}

\bibitem[{{Sokasian} {et~al.}(2002){Sokasian}, {Abel}, \&
  {Hernquist}}]{Sokasian2002}
{Sokasian}, A., {Abel}, T., \& {Hernquist}, L. 2002, \mnras, 332, 601,
  \dodoi{10.1046/j.1365-8711.2002.05291.x}

\bibitem[{{Songaila} \& {Cowie}(2002)}]{SongailaCowie2002}
{Songaila}, A., \& {Cowie}, L.~L. 2002, \aj, 123, 2183, \dodoi{10.1086/340079}

\bibitem[{{Syphers} {et~al.}(2009{\natexlab{a}}){Syphers}, {Anderson}, {Zheng},
  {Haggard}, {Meiksin}, {Schneider}, \& {York}}]{Syphers2009a}
{Syphers}, D., {Anderson}, S.~F., {Zheng}, W., {et~al.} 2009{\natexlab{a}},
  \apjs, 185, 20, \dodoi{10.1088/0067-0049/185/1/20}

\bibitem[{{Syphers} {et~al.}(2012){Syphers}, {Anderson}, {Zheng}, {Meiksin},
  {Schneider}, \& {York}}]{Syphers2012}
---. 2012, \aj, 143, 100, \dodoi{10.1088/0004-6256/143/4/100}

\bibitem[{{Syphers} \& {Shull}(2014)}]{SyphersShull2014}
{Syphers}, D., \& {Shull}, J.~M. 2014, \apj, 784, 42,
  \dodoi{10.1088/0004-637X/784/1/42}

\bibitem[{{Syphers} {et~al.}(2009{\natexlab{b}}){Syphers}, {Anderson}, {Zheng},
  {Haggard}, {Meiksin}, {Chiu}, {Hogan}, {Schneider}, \& {York}}]{Syphers2009b}
{Syphers}, D., {Anderson}, S.~F., {Zheng}, W., {et~al.} 2009{\natexlab{b}},
  \apj, 690, 1181, \dodoi{10.1088/0004-637X/690/2/1181}

\bibitem[{{The LUVOIR Team}(2019)}]{LUVOIR2019}
{The LUVOIR Team}. 2019, arXiv e-prints, arXiv:1912.06219.
\newblock \doarXiv{1912.06219}

\bibitem[{{Virtanen} {et~al.}(2020){Virtanen}, {Gommers}, {Oliphant},
  {Haberland}, {Reddy}, {Cournapeau}, {Burovski}, {Peterson}, {Weckesser},
  {Bright}, {van der Walt}, {Brett}, {Wilson}, {Millman}, {Mayorov}, {Nelson},
  {Jones}, {Kern}, {Larson}, {Carey}, {Polat}, {Feng}, {Moore}, {VanderPlas},
  {Laxalde}, {Perktold}, {Cimrman}, {Henriksen}, {Quintero}, {Harris},
  {Archibald}, {Ribeiro}, {Pedregosa}, {van Mulbregt}, \& {SciPy 1. 0
  Contributors}}]{Scipy2020}
{Virtanen}, P., {Gommers}, R., {Oliphant}, T.~E., {et~al.} 2020, Nature
  Methods, 17, 261, \dodoi{10.1038/s41592-019-0686-2}

\bibitem[{{Walther} {et~al.}(2019){Walther}, {O{\~n}orbe}, {Hennawi}, \&
  {Luki{\'c}}}]{Walther2019}
{Walther}, M., {O{\~n}orbe}, J., {Hennawi}, J.~F., \& {Luki{\'c}}, Z. 2019,
  \apj, 872, 13, \dodoi{10.3847/1538-4357/aafad1}

\bibitem[{{Worseck} {et~al.}(2019){Worseck}, {Davies}, {Hennawi}, \&
  {Prochaska}}]{Worseck2019}
{Worseck}, G., {Davies}, F.~B., {Hennawi}, J.~F., \& {Prochaska}, J.~X. 2019,
  \apj, 875, 111, \dodoi{10.3847/1538-4357/ab0fa1}

\bibitem[{{Worseck} {et~al.}(2021){Worseck}, {Khrykin}, {Hennawi}, {Prochaska},
  \& {Farina}}]{Worseck2021}
{Worseck}, G., {Khrykin}, I.~S., {Hennawi}, J.~F., {Prochaska}, J.~X., \&
  {Farina}, E.~P. 2021, \mnras, 505, 5084, \dodoi{10.1093/mnras/stab1685}

\bibitem[{{Worseck} \& {Prochaska}(2011)}]{WorseckProchaska2011}
{Worseck}, G., \& {Prochaska}, J.~X. 2011, \apj, 728, 23,
  \dodoi{10.1088/0004-637X/728/1/23}

\bibitem[{{Worseck} {et~al.}(2016){Worseck}, {Prochaska}, {Hennawi}, \&
  {McQuinn}}]{Worseck2016}
{Worseck}, G., {Prochaska}, J.~X., {Hennawi}, J.~F., \& {McQuinn}, M. 2016,
  \apj, 825, 144, \dodoi{10.3847/0004-637X/825/2/144}

\bibitem[{{Worseck} {et~al.}(2011){Worseck}, {Prochaska}, {McQuinn},
  {Dall'Aglio}, {Fechner}, {Hennawi}, {Reimers}, {Richter}, \&
  {Wisotzki}}]{Worseck2011}
{Worseck}, G., {Prochaska}, J.~X., {McQuinn}, M., {et~al.} 2011, \apjl, 733,
  L24, \dodoi{10.1088/2041-8205/733/2/L24}

\bibitem[{{Yang} {et~al.}(2020){Yang}, {Wang}, {Fan}, {Hennawi}, {Davies},
  {Yue}, {Eilers}, {Farina}, {Wu}, {Bian}, {Pacucci}, \& {Lee}}]{Yang2020}
{Yang}, J., {Wang}, F., {Fan}, X., {et~al.} 2020, \apj, 904, 26,
  \dodoi{10.3847/1538-4357/abbc1b}

\bibitem[{{Zhu} {et~al.}(2021){Zhu}, {Becker}, {Bosman}, {Keating},
  {Christenson}, {Ba{\~n}ados}, {Bian}, {Davies}, {D'Odorico}, {Eilers}, {Fan},
  {Haehnelt}, {Kulkarni}, {Pallottini}, {Qin}, {Wang}, \& {Yang}}]{Zhu2021}
{Zhu}, Y., {Becker}, G.~D., {Bosman}, S. E.~I., {et~al.} 2021, \apj, 923, 223,
  \dodoi{10.3847/1538-4357/ac26c2}

\end{thebibliography}
